\documentstyle[12pt]{article}
\begin{document}
\newcommand{\Si}{\Sigma}
\newcommand{\tr}{{\rm tr}}
\newcommand{\ad}{{\rm ad}}
\newcommand{\Ad}{{\rm Ad}}
\newcommand{\ti}[1]{\tilde{#1}}
\newcommand{\om}{\omega}
\newcommand{\Om}{\Omega}
\newcommand{\de}{\delta}
\newcommand{\al}{\alpha}
\newcommand{\te}{\theta}
\newcommand{\vth}{\vartheta}
\newcommand{\be}{\beta}
\newcommand{\la}{\lambda}
\newcommand{\La}{\Lambda}
\newcommand{\D}{\Delta}
\newcommand{\ve}{\varepsilon}
\newcommand{\ep}{\epsilon}
\newcommand{\vf}{\varphi}
\newcommand{\G}{\Gamma}
\newcommand{\ka}{\kappa}
\newcommand{\ip}{\hat{\upsilon}}
\newcommand{\Ip}{\hat{\Upsilon}}
\newcommand{\ga}{\gamma}
\newcommand{\ze}{\zeta}
\newcommand{\si}{\sigma}
\def\bfa{{\bf a}}
\def\bfb{{\bf b}}
\def\bfc{{\bf c}}
\def\bfd{{\bf d}}
\def\bfm{{\bf m}}
\def\bfn{{\bf n}}
\def\bfp{{\bf p}}
\def\bfu{{\bf u}}
\def\bfv{{\bf v}}
\def\bft{{\bf t}}
\def\bfx{{\bf x}}
\newcommand{\li}{\lim_{n\rightarrow \infty}}
\newcommand{\mat}[4]{\left(\begin{array}{cc}{#1}&{#2}\\{#3}&{#4}
\end{array}\right)}
\newcommand{\beq}[1]{\begin{equation}\label{#1}}
\newcommand{\eq}{\end{equation}}
\newcommand{\beqn}[1]{\begin{eqnarray}\label{#1}}
\newcommand{\eqn}{\end{eqnarray}}
\newcommand{\p}{\partial}
\newcommand{\di}{{\rm diag}}
\newcommand{\oh}{\frac{1}{2}}
\newcommand{\su}{{\bf su_2}}
\newcommand{\uo}{{\bf u_1}}
\newcommand{\GL}{{\rm GL}(N,{\bf C})}
\newcommand{\SL}{{\rm SL}(N,{\bf C})}
\def\sln{{\rm sl}(N,{\bf C})}
\newcommand{\gl}{gl(N,{\bf C})}
\newcommand{\PSL}{{\rm PSL}_2({\bf Z})}
\def\f1#1{\frac{1}{#1}}
\newcommand{\rar}{\rightarrow}
\newcommand{\upar}{\uparrow}
\newcommand{\sm}{\setminus}
\newcommand{\ms}{\mapsto}
\newcommand{\bp}{\bar{\partial}}
\newcommand{\bz}{\bar{z}}
\newcommand{\bA}{\bar{A}}
%\def\theequation{\thesubsection.\arabic{equation}}% the equation
               % number now does not include the section number;
               % \setcounter{equation}{0} should be put after every
               % \section{} command!!!
\newcommand{\sect}[1]{\setcounter{equation}{0}\section{#1}}
\renewcommand{\theequation}{\thesection.\arabic{equation}}
\newtheorem{predl}{Proposition}[section]
\newtheorem{defi}{Definition}[section]
\newtheorem{rem}{Remark}[section]
\newtheorem{cor}{Corollary}[section]
\newtheorem{lem}{Lemma}[section]
\newtheorem{theor}{Theorem}[section]

\vspace{0.3in}
\begin{flushright}
 ITEP-TH45/97\\
\end{flushright}
\vspace{10mm}
\begin{center}
{\Large\bf
Hierarchies of Isomonodromic Deformations\\
and Hitchin Systems}\\
\vspace{5mm}
A.M.Levin\\
{\sf Institut of Oceanology, Moscow, Russia,} \\
{\em e-mail andrl@landau.ac.ru}\\
M.A.Olshanetsky
\\
{\sf Institut of Theoretical and Experimental Physics, Moscow, Russia,}\\
{\em e-mail olshanet@heron.itep.ru}\\

\vspace{5mm}
\end{center}
\begin{abstract}
We investigate the classical limit of the Knizhnik-Zamolodchikov-Bernard
equations, considered as a system of non-stationar Schr\"{o}odinger
equations on singular curves, where times are the moduli of curves.
It has a form of reduced non-autonomous hamiltonian systems which
include as particular examples the Schlesinger equations,
Painlev\'{e} VI equation and their generalizations. In general case,
they are defined as hierarchies of isomonodromic deformations (HID)
with respect to changing the moduli of underling curves.  HID are
accompanying with the Whitham hierarchies. The phase space of HID
is the space  of flat connections of $G$ bundles with some
additional data in the marked points.  HID can be derived from some
free field theory  by the hamiltonian reduction under the action of
the gauge symmetries and subsequent factorization with respect to
diffeomorphisms of curve. This approach allows to define the
Lax equations associated with HID and the linear system whose
isomonodromic deformations are provided by HID. In addition,
it leads to description of solutions of HID by the projection method.
  In some special limit HID convert into the Hitchin systems.
In particular, for $\SL$ bundles over  elliptic curves with
a marked point we obtain in this limit the elliptic Calogero
$N$-body system.
\end{abstract}

\vspace{0.15in}
\bigskip
\title
\maketitle
\date{}

\section {Introduction}
\setcounter{equation}{0}

{\bf 1.1}
It is known for a long time that equations provided
isomonodromic deformations posses structures typical
for integrable systems, such as the Lax representation
\cite{FN,JMU}. This paper concerns the studies of isomonodromic
deformations in the spirit of
Hitchin approach to integrable systems \cite{H1}. Hitchin discovered
a wide class of hamiltonian integrable systems on
the cotangent bundles to the moduli space
 of holomorphic vector bundles over Riemann curves. Some
important facts about the Hitchin systems
 became clear later.

First of all, the Hitchin systems are related to
the Knizhnik-Zamolodchikov-Bernard (KZB) equations
\cite{KZ,B} on the critical level.
KZB equations on the critical level take the form
of second order differential equations, and their solutions
are partition functions
 of the Wess-Zumino-Witten (WZW) theory
on the corresponding Riemann curves.  It turns out that these operators
coincide with quantum second order Hitchin Hamiltonians
\cite{I1,Ne,ER,Fe,I2,BD}.
KZB eqs. for correlators of vertex operators
need to include in the construction  curves with marked points,
where vertex operators are located.
The KZB Hamiltonians  for correlators define very important classes of
quantum integrable equations such
as the  Gaudin equations (for genus zero curves), elliptic Calogero
equations  (for genus one curves), and their generalizations
\cite{FG,EK,FW}.
It means that their classical counterparts are
particular cases of Hitchin systems \cite{Ne,ER}.

Hitchin approach, based on the hamiltonian reduction of some
free field hamiltonian
theory, claims to be universal in description of integrable systems.
Essentially, it allows to present almost exhaustive
information  - integrals of motion, Lax pairs,
action-angle variables, explicit solutions via the projective method.

The main goal of this paper is to go beyond the critical
level in the classical limit
of KZB eqs. and repeat the Hitchin program as far as possible.
For generic values of level the KZB equations have the form
of non-stationer Schr\"{o}dinger equations,
where the role of times is played by the
coordinates of tangent vectors to the moduli space of curves
\cite{Fe,I2,H2}.
On the classical level they correspond to the non-autonomous hamiltonian
systems. We will call them the hierarchies
 of isomonodromic deformations (HID).
In this  situation the analog of the Hitchin phase space is
the moduli space of flat connections
${\cal A}$ over Riemann curves $\Si_{g,n}$ of genus $g$ with $n$
marked points.
While the flatness is the topological property of bundles,
 the polarization of connections ${\cal A}=(A,\bA)$ depends on
the choice of complex
structure on $\Si_{g,n}$. We consider a  bundle ${\cal P}$
 over the moduli space ${\cal M}_{g,n}$ of curves with
flat connections $(A,\bA)$
as fibers. The fibers is supplemented by elements of coadjoint
orbits ${\cal O}_a$ in the marked points $x_a$. The points of
fibers $(A,\bA)$ are
analogs of momenta and coordinates, while the base ${\cal M}_{g,n}$
serves as
a set of "times".
There exists a closed degenerated two-form $\om$
on ${\cal P}$, which is non degenerated on the fibers.
The connection $\bA$ play the same role as in the Hitchin construction,
while $A$ replaces the Higgs field. Essentially,
our construction is local - we
work over a vicinity of some fixed curve $\Si_{g,n}$ in
${\cal M}_{g,n}$.
As we already have mentioned,
the coordinates of tangent vector to ${\cal M}_{g,n}$ at $\Si_{g,n}$
play role of times, while in the Hitchin times have nothing to do
with the moduli space. The Hamiltonians of HID are
the same quadratic Hitchin
Hamiltonians, but now they are time depending.

There is some free parameter $\ka$ (the level) in our construction.
On the critical level  $(\ka=0)$, after rescaling the times, HID
convert into the Hitchin systems.
 As the later, they can be derived by the
symplectic reduction from the infinite affine space of the connections and
the Beltrami differentials with respect of gauge action on the connection.
In addition, to come to the moduli space we need the
subsequent factorization under the action
 of the diffeomorphisms of $\Si_{g,n}$,
which effectively acts on the Beltrami differentials only. Apart from
the last step, our approach is closed
to \cite{ADPW}, where the KZB systems is derived as a quantization
of the very similar symplectic quotient\footnote{Quantization of
isomonodromic deformations on rational and elliptic curves and
their relations to KZB was considered in \cite{R,Ha,Ko}}.
Due to this derivation, we find immediately the Lax pairs,
prove that the equations
of motion are consistency conditions of the isomonodromic
deformations of the
auxiliary linear problem, and, therefore, justify the notion HID.
Moreover, we describe solutions via linear procedures (the projection
method).

For genus zero our procedure leads to the Schlesinger equations.
This case was discussed earlier \cite{JMU}. We restrict
ourselves to simplest cases in which
we consider only simple poles of connections. Therefore, we don't
include in the phase space the Stokes parameters.
This phenomenon was investigated in the rational case in detail
in \cite{JMU}. Isomonodromic deformations on genus one curves
were considered in \cite{O4} and on higher genus curves in \cite{Iw}.
Here we consider
genus one curves with one marked point and obtain for $SL(2,{\bf C})$
 bundles a particular family of the Painlev\'{e} VI equations.
 Generalization of this case  on arbitrary simple groups leads
 to the multicomponent Painlev\'{e} VI related to these groups.  If we
 introduce a few marked points we come to the elliptic generalization
 of the
Schlesinger equations. In fact, the concrete systems we consider here
are the deformation from the critical level of those Hitchin systems
that described in \cite{Ne}.
\bigskip

{\bf 1.2. Painlev\'{e} VI and Calogero equations}
The very instructive (but not generic) example
of our systems is the Painlev\'{e} VI
equation. It depends on four free parameters
$(PVI_{\al,\be,\ga,\de})$ and has the
form
$$
\frac{d^2X}{dt^2}=\frac{1}{2}(\frac{1}{X}+\frac{1}{X-1}+\frac{1}{X-t})
(\frac{dX}{dt})^2-
(\frac{1}{t}+\frac{1}{t-1}+\frac{1}{X-t})\frac{dX}{dt}+
$$
\beq{I.1}
+\frac{X(X-1)(X-t)}{t^2(t-1)^2}(\al+\be\frac{t}{X^2}+
\ga\frac{t-1}{(X-1)^2}
+\de\frac{t(t-1)}{(X-t)^2}).
\eq
$PVI_{\al,\be,\ga,\de}$  has a lot
of different applications (see, for example \cite{PT}).
It is a hamiltonian systems \cite{O1}. We will write
the symplectic form and the Hamiltonian below in another variables.
This equation was derived firstly by Gambier \cite{G}, as an equations
which has not solutions with moveable singularities.
Among  distinguish features of this equation we are interesting in
its relation to the isomonodromic deformations
of linear differential equations.
This approach was investigated  by Fuchs \cite{F}.

There exsits elliptic form of $PVI_{\al,\be,\ga,\de}$ derived by
Painlev\'{e} itself \cite{Pa}. It was investigated recently by
 Manin \cite{Ma} in connection with Frobenius varieties.
It sheds light on connection of the Painlev\'{e} equations with the
Hichin systems,
 and, thereby,
with the KZB equations. We present shortly this approach.

Let $\wp(u|\tau)$ be the Weierstrass function on the elliptic curve
$T^2_{\tau}={\bf C}/({\bf Z}+{\bf Z}\tau)$, and
$$e_i=\wp(\frac{T_i}{2}|\tau),~~(T_0,\ldots,T_3)=(0,1,\tau,1+\tau).$$
Consider instead of $(X,t)$ in (\ref{I.1})  the new variables
\beq{I.2}
(u,\tau)\rar (X=\frac{\wp(u|\tau)-e_1}{e_2-e_1},t=\frac{e_3-e_1}{e_2-e_1}).
\eq
Then $PVI_{\al,\be,\ga,\de}$ takes the form
$$
\frac{d^2u}{d\tau^2}=-\p_uU(u|\tau),
$$
\beq{I.3}
U(u|\tau)=\frac{1}{(2\pi i)^2}\sum_{j=0}^3\al_j\wp(u+\frac{T_j}{2}|\tau),
\eq
$$
(\al_0,\ldots,\al_3)=(\al,-\be,\ga,\f1{2}-\de).
$$
 The hamiltonian form of (\ref{I.3}) is defined
by the standard symplectic form
\beq{I.4}
\om_0=\de v\de u,
\eq
and the Hamiltonian
\beq{I.5}
H=\frac{v^2}{2}+U(u|\tau).
\eq
The equation of motion (\ref{I.3}) can be derived from the action $S$
\beq{I.5a}
\de S=v\de u-H\de\tau.
\eq
We can consider (\ref{I.4}),(\ref{I.5}) as non-autonomous hamiltonian system
with \\
the time-dependent potential.

To find symmetries we consider two-form
\beq{I.5b}
\om=\om_0-\de H\de\tau=\de v\de u-\de H\de\tau.
\eq
The semidirect product of ${\bf Z}+{\bf Z}\tau$ and the modular
group acting on the dynamical variables $(v,u,\tau)$ are the symmetry
of (\ref{I.5b}). We consider them in detail in Sect.7.

Let us introduce the new parameter $\ka$ and instead of (\ref{I.5a}) consider
\beq{I.6}
\om=\om_0-\f1{\ka}\de H\de\tau.
\eq
Then, (\ref{1.3}) takes the form
\beq{I.3a}
\ka^2\frac{d^2u}{d\tau^2}=-\p_uU(u|\tau).
\eq
It corresponds to the overall rescaling of constants
$\al_j\rar\frac{\al_j}{\ka^2}$.
Put $\tau=\tau_0+\ka t^H$ and consider the system in the limit
 $\ka\rar 0$.
We come to the equation
\beq{I.7}
\frac{d^2u}{(dt^H)^2}=-\p_uU(u|\tau_0),
\eq
It is just the rank one elliptic Calogero-Inozemtsev equation
\cite{Ca,In,TV},
which we denote $CI_{\al,\be,\ga,\de}$. Thus, we have in this limit
\beq{I.7a}
PVI_{\al,\be,\ga,\de}\stackrel{\ka\rar 0}
{\longrightarrow}CI_{\al,\be,\ga,\de}.
\eq

So far we don't now how to manage with the general forms of both types
of equations for arbitrary values of constants.
Here we consider only  the one-parametric family in (\ref{I.3a})
$PVI_{\frac{\nu^2}{4},-\frac{\nu^2}{4},
\frac{\nu^2}{4},\f1{2}-\frac{\nu^2}{4}}$
$$\al_j=\frac{\nu^2}{4},~~(\al=\frac{\nu^2}{4},\be=-\frac{\nu^2}{4},
\ga=\frac{\nu^2}{4},\de=\f1{2}-\frac{\nu^2}{4}).$$
The potential (\ref{I.3a}) takes the form
\beq{I.8}
U(u|\tau)=\frac{1}{(2\pi i)^2}\nu^2\wp(2u|\tau),
\eq

We will prove that (\ref{I.4})(\ref{I.5}) with the potential
(\ref{I.8}) describe the dynamic of flat connections
of ${{\rm SL}(2,{\bf C})}$  bundles over elliptic curves $T_{\tau}$
with one marked
point $\Si_{1,1}$. In fact, $u$ lies on the Jacobian of $T_{\tau}$, $(v,u)$
are related to the flat connections, and $\tau$ defines a point
in ${\cal M}_{1,1}$. Roughly speaking,
the triple $(v,u,\tau)$ are the coordinates in the total space
of the bundle ${\cal P}$ over ${\cal M}_{1,1}$,
$\om$ (\ref{I.6}) is the two-form on ${\cal P}$, and $\om_0$ is the
symplectic form on its fibers. This finite-dimensional Hamiltonian
system is derived as a  quotient of  infinite-dimensional
phase space of $(A,\bA)$ connections and
the Beltrami differentials as times under the action of gauge transforms
and diffeomorphisms of $\Si_{1,1}$. This approach directly
leads to
 the Lax linear system and allows to define
 solutions of the Cauchy problem via the projection procedure.
Simultaneously, we describe the auxiliary linear problem whose
isomonodromic deformations are governed by
this particular family of the Painlev\'{e} VI. The discrete symmetries of
(\ref{I.5a}) are nothing else as the remnant gauge symmetries.
On the critical level it is
just two-body elliptic Calogero system. The corresponding quantum
system is identified with the KZB equation for the one-vertex correlator.
In the similar way $PVI_{\frac{\nu^2}{4},-\frac{\nu^2}{4},
\frac{\nu^2}{4},\f1{2}-\frac{\nu^2}{4}}$ is the classical limit of the
KZB for $\ka\neq 0$. This example will be analyzed in detailed in Sect.7.

Solutions of $PVI_{\frac{\nu^2}{4},-\frac{\nu^2}{4},
\frac{\nu^2}{4},\f1{2}-\frac{\nu^2}{4}}$ in
particular case $(\nu^2=\oh)$ were found by Hitchin
\cite{H3} in connection with his investigations of anti-self-dual
 Einstein metrics.
They are written in terms of theta functions. But for generic values of
$\nu$ solutions are genuine  Painlev\'{e} transcendents \cite{O3}.

\bigskip

{\bf 1.3. Whitham equations}
The previous example is not entirely exhaustive. A special phenomena occurs
 if $\dim{\cal M}_{g,n}>1$. We have as many Hamiltonians as
$\dim{\cal M}_{g,n}$ - each Hamiltonian $H_s$ is attached
to the tangent vector $t_s$
to ${\cal M}_{g,n}$ at some fixed point $\Si_{g,n}$.
 There are consistency conditions of the equations of motion
for the non-autonomous multi-time hamiltonian systems. They take a form
of the classical zero-curvature conditions for the connections
\beq{I.9}
\p_{t_s}+H_s,
\eq
 where the commutator is replaced by the Poisson
brackets. The later is the inverse tensor to the symplectic form, defined
on the fibers. The flatness conditions is the so-called the Whitham hierarchy,
(WH)
defined in the same terms by Krichever \cite{Kr1}, though the starting point
of his construction is different and based on the averaging procedure.
Let $S$ be the action for HID $(\de S=\de^{-1}\om$
as in(\ref{I.5a})). The
$\tau$-function of WH is
$$\log\tau=S.$$
It allows to find the Hamiltonians. For HID in the rational case
the $\tau$-functions
were investigated in \cite{JMU}. For the Painlev\'{e} I-VI
they were considered in \cite{O2}.

The quantum analog of the Whitham equations was exploited in \cite{H2,Fe}
to construct the KZB connections. The quantum version of (\ref{I.9}) is an
operator, acting in the space $V$ of sections of holomorphic line
bundle over moduli of flat connections. The quantum WH is the flatness
of the bundle of projective spaces ${\bf P}V$ over the Teichm\"{u}ller
space ${\cal T}_{g,n}$.

\bigskip

{\bf 1.4. Outline}
We start in sect. 2 with a general setup about flat bundles over
singular curves.
In sect. 3 we present the basic facts about the abstract nonautonomous
hamiltonian equations. They defined by a degenerate closed two-form,
on the extended phase space, which is a bundle over the space of times.
The Whitham equations  occur on this stage.
We remind the symplectic reductions technique, which is applicable  in
degenerate case as well.
Sect. 4 contains our  main result
- the derivation of the HID, corresponding to the flat bundles
over Riemann curves.  In sect. 5 we discuss two limits
- the level zero limit of HID to the
Hitchin systems and the classical limit of
general KZB equations  to HID. Then we analyze in detail
two feasible examples
- flat bundles over  rational curves, leading to the Schlesinger
equations (sect. 6), and flat bundles over elliptic curves,
responsible for the Painlev\'{e} VI type equations, and the elliptic
 Schlesinger equations (sect. 7).  In particular, we apply the projection method to construct perturbatively solutions of the Schlesinger equations.
In the Appendix we summarize
some results about the elliptic functions, which are used in sect. 7.

{\bf Acknowledgments}\\
{\sl We would like to thank our colleges  V.Fock, A.Losev and N.Nekrasov
for fruitful and illuminating discussions.
 We are grateful to the Max-Planck-Institut f\"{u}r Mathemamatik in Bonn
for the hospitality, where this work was started. We are
thankful to Yu.Manin - his lectures and discussions with
him concerning PVI, stimulated
our interests to these problems. The work is
supported in part by grants INTAS 94-4720, Award No.
RM1-265 of the Civilian Research \& Development Foundation
(CRDF) for the Independent States of the Former Soviet Union (A.L);
grants INTAS 96-538, RFBR-96-02-18046, Award No.
RM2-150 of the Civilian Research \& Development Foundation
(CRDF) for the Independent States of the Former Soviet Union, INTAS
(M.O.);  grant  96-15-96455  for support of scientific schools
(A.L., M.O).}

\section {Flat bundles over singular curves}
\setcounter{equation}{0}

We will describe the general setup more or less naively. We don't
consider in this section the mechanism of symplectic reduction
in detail and postpone it on Sect.4
We will consider three cases:
1) smooth proper (compact) algebraic curves;
2) smooth proper algebraic curves with punctures;
3) proper algebraic curves with nodal singularities (double points).

Let $G$ be a semisimple group and $V$
be its exact representation; f.e. $G=\SL$ and
 $V$ is a $N$-dimensional vector space with a volume form.
\bigskip

{\bf 2.1. Smooth curves.}

Let $S$ be a smooth oriented compact surface of genus $g$. Let us consider
the moduli space $FBun_{S,G}$ of flat $V$-bundles on $S$. This space can be
considered
as the quotient of the space $FConn$ of flat $C^{\infty}$
connections on trivial $V$-bundle
by action of the gauge group $\cal G$ of $G$-valued
$C^{\infty}$-functions on $S$, or as a result of
hamiltonian reduction of space $Conn$ of all connections by action of the
gauge group. The symplectic form on the space $Conn$ is the form
\beq{G.1}
\om=\int_{S}<\delta {\cal A}, \delta{\cal A}>,
\eq
 where $<\cdot,\cdot>$ denotes the Killing form,
$\delta{\cal A}$ is a Lie$(G)$-valued one-form on $S$. So,
$<\delta{\cal A}, \delta{\cal A}>$  is a two-form,
and the integral is well defined.
The  dual space to the gauge algebra Lie is the space of
Lie$(G)$-valued
two-forms on $S$, and the momentum map corresponds
to any connection $\cal A$
its curvature $F_{\cal A}= d {\cal A} +\oh [{\cal  A},{\cal A}]$. Hence,
the preimage of $0$ under momentum map is the space of flat connections.

A complex structure $\Si$ on $S$ is a differential operator
$\overline{\partial}:{\Om}_{C^\infty}^0\to{\Om}_{C^\infty}^1$;
its kernel is the space
of holomorphic functions. For any connection we can consider
$\overline{\partial}$-part of connection; it defines a holomorphic bundle.
A section is holomorphic, if it belongs to the kernel of this operator.
Let us recall that:\\
1. Connection on the holomorphic bundle ${\cal E}$ is the operator of type
$\partial +A$.\\
2. Connection is holomorphic if this operator commutes with
operator of complex structure on the bundle.\\
3. Connection is flat if its square vanishes; for curves this condition
is empty.\\
4. The space of all holomorphic connections on ${\cal E}$ is an infinite
dimensional affine space and the gauge group ${\cal G}$ acts on it
by affine transformations.\\
5. The quotient of stable holomorphic connections with respect to
the gauge transformations is the
 moduli space  $Bun_{\Si,G}$ of holomorphic vector bundles on $\Si$.\\

Let $b$ be a point in  $Bun_{\Si,G}$ and ${\cal V}_b$ is the corresponding
$V$-bundle on $\Si$. Denote by $ad_b$ the bundle of endomorphisms
of ${\cal V}_b$ as bundle of Lie algebras and  by $Flat_b$ the space of
flat holomorphic connections on holomorphic bundle ${\cal V}_b$.

Consider the map from $FBun_{S,G}$ onto the moduli space  $Bun_{\Si,G}$.
The fiber of the projection   $FBun_{S,G}\to Bun_{\Si,G}$ over a point $b$
is naturally isomorphic to $Flat_b$. These fibers
are Lagrangian with respect
to $\om$ (\ref{G.1}) - any complex structure on $S$ defines a
polarization  on  $FBun_{S,G}$.

 The tangent space to $Bun_{\Si,G}$  at the point
$b$ is canonically isomorphic to the first cohomology group
$H^1(\Si, ad_b)$.
The space  $Flat_b$ of holomorphic connections on ${\cal V}_b$ is an
affine space over the vector space
$H^0(\Si,ad_b\otimes \Omega ^1)$ of holomorphic
$ad_b$-valued one-forms, since a difference between any connections is
a $ad_b$-valued differential form.
The vector spaces $H^1(\Si, ad_b)$ and $H^0(\Si,ad_b\otimes \Omega ^1)$
are dual.

The Dolbeault representation of cohomology classes corresponds to the
description based on the hamiltonian reduction. Indeed, let us decompose
$\cal A$ into $(1,0)$ and $(0,1)$ parts:
${\cal A}=A+\overline{A}$. From the zero curvature condition
we get that $(1,0)$-part $\delta A$ must be $\overline{A}$-holomorphic:
$\overline{\partial}_{ A}\delta A\equiv
(\overline{\partial}+\overline{A})\delta A=0$, and
$\delta\overline{A}$
is defined up to the infinitesimal gauge transformations
$\delta\overline{A}\to \delta\overline{A}
+\overline{\partial}_{A} h$.

\bigskip
{\bf 2.2. Curves with punctures}

For any two isomorphic representations $V$ and $V'$ of $G$ denote by
$Isom(V, V')$ the space of $G$-isomorphisms between them; this space
is a principal homogeneous space over $G$.
For a curve $\Si_n$ with $n$ marked points $x_j$ we replace the moduli
space $Bun_{\Si,G}$ by moduli space $Bun_{\Si,{\bf x},G}$,
$({\bf x}=(x_1,\ldots,x_n))$ of holomorphic
$V$-bundles $\cal V$ with the trivializations $g_j:{\cal V}_{x_j}\to V$
of fibers at the marked points. We have natural ``forgetting''
projection $\pi :Bun_{\Si,{\bf x},G}\to Bun_{\Si,G}$.
The fiber of this projection is the  product
$\prod Isom ({\cal V}_{x_j},V)$
of the spaces of isomorphisms between fibers of bundle at marked points
and $V$. The projection $\pi$ can be treated as a reduction by
action of $n$ copies of the group $G$ which acts transitively on
the fibers
$$
Bun_{\Si,G}=Bun_{\Si,{\bf x},G}/\prod G.
$$
The bundle of endomorphisms of this data is the bundle $ad_b(-{\bf x})$
of endomorphisms, vanishing at the marked points,
since non vanishing endomorphisms
change trivializations. Hence, the tangent space to
$Bun_{\Si,{\bf x},G}$
is isomorphic to $H^1(\Si, ad_b(-{\bf x}))$. The dual space to
this space is  $H^0(\Si, ad_b({\bf x})\otimes \Omega ^1)$.
Consequently, in order to get the symplectic variety
we must replace the affine space $Flat_b$ of flat
holomorphic connection by the affine space $Flat_b(\log)$ of
flat connections with logarithmic singularities.
As a result, we get the moduli space $FBun_{\Si,{\bf x},G}$ of
triples: (holomorphic bundle, trivializations at marked points,
holomorphic connection with logarithmic singularities). Again,
we have the Lagrangian projection
$FBun_{\Si,{\bf x},G}\to Bun_{\Si,{\bf x},G}$.
The fiber of this projection is an affine space over
 $H^0(\Si, ad_b({\bf x})\otimes \Omega ^1)$.

The product of $n$ copies of $G$ acts on $FBun_{X,\bfx,G}$ by
changing trivialization of fibers, this action is hamiltonian.
We can consider the hamiltonian reduction with respect to
this action. As a result, for any collection of $G$-orbits of adjoint
action we get the moduli space
$FBun_{\Si,{\bf x},G,\{{\cal O}_j\}}$ of pairs (holomorphic bundle,
holomorphic connection with logarithmic singularities with residues
in orbits ${\cal O}_j$).

According to the Dolbeault theorem, the tangent space
$H^1(\Si,ad_b(-{\bf x}))$ can be realized as
the space of $(0,1)$ $ad_b$-valued forms {\em holomorphically}
vanishing at the marked points modulo $\overline{\partial}_A$-coboundaries
of  {\em holomorphically}
vanishing at marked points $ad_b$-valued functions. A function (or
one-form) is referred to be holomorphically vanishing at the point $x$,
if it has
asymptotics $(z-x)O(1)$ at this point.

The second description: the tangent space $H^1(\Si,ad_b(-{\bf x}))$
is isomorphic to the space of all $(0,1)$ $ad_b$-valued
forms modulo $\overline{\partial}_A$-coboundaries
of vanishing at marked points $ad_b$-valued functions.
Indeed, we have a natural embedding of the space of holomorphically
vanishing forms to the space of all forms and the space of holomorphically
vanishing functions to the space of vanishing functions. It defines
the map $\phi$ from the space from first description to space from second
description. We can assume that $\overline A$ vanishes
at  small vicinities
of marked points. Then $\overline{\partial}_{\overline A}$ equals
$\overline{\partial}$ at these vicinities. Let $h$ be a function,
 vanishing at the marked points which is not holomorphically vanishing,
 $$
h\equiv \sum_{i=1}^{\infty} a_i\overline{(z-x_j)}^i+(z-x_j)o(1).
$$
Then
$$
\overline{\partial}h
\sim(\sum_{i=1}^{\infty}i
a_i\overline{(z-x_j)}^{(i-1)}+(z-x_j)o(1))d\,\overline{z}
$$
is not
holomorphically vanishes. So, the map $\phi$ is injective.
At the other hand,
for any form $\nu$ with asymptotics
$$
\nu\sim (\sum_{i=0}^{\infty}
a_i\overline{(z-x_j)}^i+(z-x_j)o(1))d\,\overline{z}
$$
 at the points $x_j$ we can choose some function $\phi$ such that
$\nu -\overline{\partial}_{A}\phi$ has asymptotics $(z-x_j)o(1)$,
since for any collection of asymptotics
$(\sum_{i=o}^{\infty}\frac{1}{i+1}a_i
\overline{(z-x_j)}^{i+1}+(z-x_j)o(1))$
a function with such asymptotic exists.
 So, the map $\phi$ is surjective.

The third equivalent description of this space is defined as
the cokernel of the map:
$$
\Om_{C^\infty}^{0,0}(ad_b)\to  \Om_{C^\infty}^{0,1}(ad_b)+
\sum ad_b|_{x_i};~
h\to (\overline{\partial}_{A} h, h(x_i)).
$$
Indeed, for any element $ (\delta\overline{A}, \{a_i\})$ in  the cokernel
we can choose its representative with vanishing second part
$ \{a_i\}$. This choice is unique up to
$\overline{\partial}_{A} h$ for vanishing at the marked points functions
$h$. This description is adapted to the forgetting map $\pi$.
The local part $\{a_i\}$ corresponds to the tangent space of fiber
and the global part $\delta\overline{A}$ corresponds to the tangent space
of base.

By ``integration'' of the action of the gauge Lie algebra $ad_b$ to
the action of the gauge Lie group $\cal G$ we get two description
of the moduli space.

A. The moduli space  $Bun_{\Si,{\bf x},G}$ is the quotient of the
space of $\overline{\partial}$-connections $\overline A$ on
the trivial $V$-bundle by action of
the reduced gauge group of $G$-valued functions, whose values
at the marked points are equal to the neutral element $1$ of the group
$G$. The isomorphisms of fibers with $V$ are identical maps.

B. The moduli space  $Bun_{\Si,{\bf x},G}$ is the quotient of the
space of pairs (a $\overline{\partial}$-connection $\overline A$ on
the trivial $V$-bundle,
a collection $g_j$ of elements of $G$) by action of the gauge group
$\cal G$. The isomorphisms of fibers with $V$ are $g_i$.
The isomorphism of these descriptions can be proved by the following
consideration. For any collection ($\overline A$, $g_j$) we can choose
the gauge equivalent collection with $g_j=1$, and such collections are
equivalent up to action of the reduced gauge group.

In what
follows we assume the second description of $Bun_{X\Si,{\bf x},G}$

Let us consider the space of data ($\overline{\partial}$-connections
$\overline A$ on the trivial $V$-bundle, a collection $g_j$ of
elements of $G$, $\partial$-connection $A$ with logarithmic
singularities at marked points). This space is symplectic with the
form
$FBun_{X,\{x_j\},G}$ equals to Hamiltonian reduction of this space
by action of gauge group $\cal G$.

It is worthwhile to note that (in contrast with Case 1) this construction
is essentially based on the complex structure on the
surface, since $\overline A$ is nonsingular and $A$ has singularities at
the marked points. We can consider the moduli space of flat bundles over
noncompact surface $S\setminus \bigcup x_j$ = $G$-representations of
fundamental group of $S\setminus \bigcup x_j$. According to
Deligne, for any complex structure on $S$ any stable representation
can be realized by connections with logarithmic singularities on
holomorphic bundle. Hence, any connection $\cal A$ is gauge
equivalent to a connection with regular $\overline{\partial}$-part,
but corresponding gauge transformation has any singularities at
marked points a priori. This fact makes such approach very complicated.

\bigskip

{\bf 2.3. Curves with double points}

A curve with a double point (the nodal singularity)
 can be treated as a ``limit'' of
nonsingular curves under pinching of some circle. If this circle
is homological to zero, then the resulting curve is the union of two
intersecting smooth curves; and sum of genera of these curves
is equal to the genus of nonsingular curves. If this circle is
 homologically nontrivial, then the singular curve is a smooth curve with
``glued'' two different points. The genus of this curve is less
by one that genus of initial curves. The normalization of
singular curve (``disglueing'' of singularity) is a smooth curve (not
connected, in general) with marked points.

Let us fix some notations. Denote by ${\cal M}_g$ the moduli
space of smooth curves of genus $g$, and denote by ${\cal
M}_{g,n}$   the moduli
space of smooth curves of genus $g$ with $n$ different marked
points (${\cal M}_{g}={\cal M}_{g,0}$) .  Let
$\overline{\cal M}_{g,n}$ be the Deligne-Mumford
compactification of ${\cal
M}_{g,n}$. Then the compactification divisor $D_{\infty}=\overline{\cal
M}_{g}\setminus {\cal M}_{g}$ is the union of components
$D_{\infty}^{g_1,g_2}$, $g_1+g_2=g$ and $D_{\infty}^{g-1}$.
These components are covered by
$\overline{\cal M}_{g_1,1}\times\overline{\cal M}_{g_2,1}$
and $\overline{\cal M}_{g-1,2}$ correspondingly.

Consider the moduli space $Bun_{g,\Si}$ of pairs (smooth curve of
genus $g$, $G$-bundle on it). Evidently, this space is fibered
over ${\cal M}_{g}$ with $Bun_{X,G}$ as fibers.
 Unfortunately, we don't now the canonical compactification
  $\overline{Bun_{g,X}}$, which is fibered over $\overline{\cal
M}_{g}$. Let us assume that such compactification does exist.
Then the open part of fiber over singular curve must be
described as moduli space of the following data
($G$-bundle over normalized curve, isomorphisms  of fibers over
``glued'' points).

 Denote by $\Si^0$ a singular curve with nodal points
$y_1,y_2\cdots y_n$, and denote by $\Si$ its normalization, $x_a$ and
$x_{n+a}$ are preimages of $y_a$ under normalization map.
 Then moduli space $Bun_{\Si^0,G}$, corresponding to $\Si^0$
is the moduli space
of (holomorphic $V$-bundle on $\Si$, isomorphisms between fibers
of this bundle over $x_a$ and $x_{n+a}$).
This space is quotient of $Bun_{\Si,{\bf x},G}$ by action of $n$-copies
of group $G$, $j$-th $G$ acts on $Isom({\cal V}_{x_a}, V)$ and
$Isom({\cal V}_{x_{n+a}}, V)$ from the right:
$$
\{h_{a}\}_{1\leq a\leq
n } :\{g_{a},g_{a+n}\}_{1\leq a\leq n}\to
\{g_{a}h_{a},g_{a+n}h_{a}\}_{1\leq a\leq n}.
$$
Corresponding symplectic variety $FBun_{\Si^0,G}$ is the result
of symplectic reduction
of\\
  $FBun_{\Si,{\bf x},G}$ by the described above action with zero
level of momentum map. This space is the moduli space of data
(holomorphic $V$-bundle on $X$, isomorphisms between fibers of
this bundle over $x_a$ and $x_{n+a}$, connection with logarithmic
singularities such that residue at points $x_a$ and $x_{n+a}$
are opposite). In the last part of data we use isomorphisms of fibers
from the second part of data.

\section{Hamiltonian formalism}
\setcounter{equation}{0}

We  consider nonautonomous hamiltonian systems. For this type of systems
it is the custom in the classical mechanics to deal with
a degenerate symplectic form on the extended phase space,
which includes beside the usual coordinates and
momenta, the space of times and corresponding to
them Hamiltonians as the conjugate
variables. Then  the hamiltonian equations
of motion are defined as variations
 of dynamical variables along the null leafs of this symplectic form.
\bigskip

{\bf 3.1. Equations of motion}

Let ${\cal R}$ be a
(infinite-dimensional) phase space endowed with the non-degenerate
 symplectic structure. For simplicity,
we take it in the canonical form
$$ \om_0=(\de {\bf v},\de {\bf u}),$$
$${\bf v}=(v_1,\ldots,v_i\ldots),~~{\bf u}=(u_1,\ldots,u_i\ldots).$$
Consider the space of "times"
${\cal N}=\{\bft=(t_1,\ldots,t_a,\ldots)\}$
 and corresponding dual to
 times Hamiltonians $(H^1,\ldots,H^a,\ldots)$  on ${\cal R}$
depending on times as well.
We consider the extended phase space which is the bundle
${\cal P}$ over ${\cal N}$ with fibers ${\cal R}$.
Introduce a symplectic form on ${\cal P}$
\beq{01}
\om=\om_0-\sum_a \de H^a\de t_a=(\de {\bf v},\de {\bf u})-
\sum_a \de H^a\de t_a.
\eq
This form is closed but degenerated on ${\cal P}$.  The vector fields
$${\cal V}^a=\sum_i(A^a_i\p_{v_i}+B^a_i\p_{u_i})+\p_{t_a}.$$
 lie in the kernel of $\om$ iff
$$
A^a_i+\frac{\p H^a}{\p u_i}=0,~~
-B^a_i+\frac{\p H^a}{\p v_i}=0,
$$
\beq{H1}
-A^a_i\frac{\p H^b}{\p v_i}-B^a_i\frac{\p H^b}{\p u_i}-
\frac{\p H^a}{\p t_b}+
\frac{\p H^b}{\p t_a}=0.
\eq
Then the vector field ${\cal V}^a\in{\rm Ker} ~\om $ takes the form
$$
{\cal V}^a=-\frac{\p H^a}{\p u_i}\p_{v_i}+
\frac
{\p H^a}{\p v_i}\p_{u_i}+
\p_{t_a}.$$
It can be checked immediately that they commute.
For any functions $f(V,U,t)$ on ${\cal P}$ its evolution is defined as
\beq{H2}
\frac{df({\bf v},{\bf u},{\bf t})}{dt_a}=
{\cal V}^af({\bf v},{\bf u},{\bf t}),
~~~{\cal V}^a\in{\rm Ker} ~\om.
\eq
It follows  from (\ref{H1}) that the Hamiltonians subject
to the classical zero curvature conditions
({\sl the generalized Whithem hierarchy})
\beq{H3}
\frac{d H^a}{dt_b}-\frac{dH^b}{dt_a}+
\{H^b,H^a\}_{\om_0}=0,~a,b=1\ldots
\eq
Evidently, for the time independent Hamiltonians (\ref{H2}),(\ref{H3})
give the standard approach.
Define the action $S$ of the system as $\de^{-1}\om=\de S$
\beq{S}
\de S=({\bf v},\de {\bf u})-\sum_a H_a\de t_a
\eq
and the $\tau$-{\sl function}
\beq{tau}
\de\log\tau=\de S.
\eq
 The equations of motion
 can be written down in the Hamilton-Jacobi form
$$
\frac{\p S}{\p t_a}=-H_a(\frac
{\de S}{\de {\bf u}},{\bf u},{\bf t}).
$$
\bigskip

{\bf 3.2. Moment map}

The symplectic reduction
can be applied to the nonautonomous
system with symmetries in  the standard way.
We shortly repeat this approach.
Let ${\cal G}$ be a symmetry group of the system. It means that for any
$\ep\in$Lie$({\cal G})$ there exists the vector field acting on ${\cal P}$
such that
the Lie derivative ${\cal L}_\ep$ annihilates $\om$ (\ref{01})
$$
{\cal L}_\ep\om=(\de j_\ep+j_\ep\de)\om=0.
$$
The vector field is called a {\it hamiltonian
vector field} with respect to $\om$.
Since $\om$ is closed we can write locally
$j_\ep\om=\de F_\ep$. If $j_\ep\om$
is exact then $j_\ep$ is the {\it strictly hamiltonian vector field}.
The function
$F_\ep=F_\ep({\bf v},{\bf u},{\bf t})$ is a linear function on
Lie$({\cal G})$
$$
F_\ep=<\ep,{\cal J}({\bf v},{\bf u},{\bf t})>
$$
 and thereby defines the {\it the moment map }
$$
{\cal J}:{\cal P}\rar Lie^*({\cal G}).$$
Assume that we put the moment constraint
\beq{H4}
{\cal J}({\bf v},{\bf u},{\bf t})=0.
\eq
The quotient
$${\cal P}^{\rm red}={\cal P}//{\cal G}:={\cal J}^{-1}(0)/{\cal G}$$
is a symplectic space with the symplectic form $\om^{\rm red}$
which is the reduction of the original form $\om$
(\ref{01}). It is defined by
the two step procedure.\\
i) Fixing the gauge. In other words, there should be defined a surface
in the phase space ${\cal P}$, which is transversal to the orbits
of the gauge group ${\cal G}$. \\
ii) Solving the moment constraint equations (\ref{H4}) for
the dynamical variables, restricted to the gauge fixing surface.

The main difference with the autonomous hamiltonian systems is
that we demand the gauge invariance for the whole degenerate
symplectic form (\ref{01}) only and don't consider the form $\om_0$
and the Hamiltonians separately.

\section{Symplectic reduction and factorization}
\setcounter{equation}{0}

{\bf 4.1. Space of "times" ${\cal N}'$ }.

 Let $\Si_{g,n}$ be
a Riemann curve  of genus $g$ with
$n$ marked points . Let us fix the complex structure
on $\Si_{g,n}$ by defining
local coordinates $(z,\bz)$ in open maps covering $\Si_{g,n}$.
Assume that the marked points $(x_1,\ldots,x_n)$
are in the generic position, i.e. there exists a set of their vicinities
$({\cal U}_1,\ldots,{\cal U}_n)$,
such that ${\cal U}_a\cap{\cal U}_b=\emptyset$ for $a\neq b$.

The deformations
of the basic complex structure are determined by the
of  Beltrami differentials $\mu$,
which are smooth $(-1,1)$ differentials on $\Si_{g,n}$.
We identify this set with the space of times ${\cal N}'$.
 The Beltrami  differentials  can be defined in the following way.
Consider a chiral smooth transformation of $\Si_{g,n}$, which in
some local map can be represented as
\beq{2a.2}
w=z-\ep(z,\bz),~~\bar{w}=\bz
\eq
Up to the conformal factor $1-\p \ep(z,\bz)$
the corresponding one-form $dw$  is equal
\beq{1a.2}
dw=dz-\mu d\bz,~~\mu=\frac{\bp \ep(z,\bz)}{1-\p \ep(z,\bz)}.
\eq
The Beltrami differential defines the new holomorphic structure -
the deformed antiholomorphic operator annihilating $dw$, while
the antiholomorphic structure is kept unchanged
$$
\p_{\bar{w}}=\bp+\mu\p,~~\p_w=\p.
$$
In addition, assume that $\mu$ vanishes
in the marked points
\beq{1b.2}
\mu(z,\bz)|_{x_a}=0.
\eq

In our construction we consider small deformations of the basic complex
structure $(z,\bz)$. It allows  to replace (\ref{1a.2}) by
\beq{2b.2}
\mu=\bp \ep(z,\bz).
\eq
Nevertheless, in some cases we will use the exact representation
(\ref{1a.2}) as well.
\bigskip

{\bf 4.2 Fibers} ${\cal R}'$.

 Let ${\cal E}$ be a principle stable
$G$ bundle over a Riemann curve $\Si_{g,n}$ . Assume that $G$
is a complex simple Lie group .
The phase space ${\cal R}'$ is constructed by the following data:\\
i)the affine space $\{{\cal A}\}$ of Lie$(G)$-valued connection
 on ${\cal E}$.\\
It has the following component description:\\
a)   $C^{\infty}$ connection $\{\bar{A}\}$ ,
corresponding to the $d\bar{w}=d\bz$ component of ${\cal A}$;\\
b)The dual  to the previous space the space $\{A\}$ of $dw$  components
of connection ${\cal A}$. $A$ can have simple poles in the marked points.
Moreover, assume that $A\mu$ is a well defined function ;\\
ii)cotangent bundles
 $T^*G_a=\{(p_a,g_a),~p_a\in {\rm Lie}^*(G_a),
~g_a\in G_a\},~(a=1,\ldots,n)$ in the points
$(x_1,\ldots,x_n)$.\\
There is the canonical  symplectic form on ${\cal R}'$
\beq{1.2}
\om_0=\int_{\Si}<\de A,\de\bar{A}>+
2\pi i\sum_{a=1}^n\de<p_a,g_a^{-1}\de g_a>,
\eq
where $<~,~>$ denotes the Killing form on Lie$(G)$.

According with Sect.2 we can consider the elements  $g_a$
as trivializations of fibers in the marked
points and $p_a$ as the residues of holomorphic connections in the same
points
$$g_a\in Isom({\cal V}_a,V),~(V\sim {\rm Lie}(G)),~~
p_a=res|_{x_a}A.
$$
 Thus, the symplectic form (\ref{1.2}) is the generalization of
(\ref{G.1}) on the singular curves.
\bigskip

{\bf 4.3. Extended phase space} ${\cal P}'$.

According to the general prescription
the bundle ${\cal P}'$ over ${\cal N}'$ with
${\cal R}'$ as the fibers plays role of the extended phase space.
Consider the degenerate form  on  ${\cal P}'$
\beq{2.2}
\om=\om_0-
\frac{1}{\ka}
\int_{\Si}<\de A, A>\de \mu.
\eq
Thus, we deal with the infinite set of Hamiltonians $< A, A>(z,\bz)$,
parametrized by points of $\Si_{g,n}$ and corresponding
set of times $\mu(z,\bz)$.
We apply the formalism presented in the previous
section to these systems.
\bigskip

{\bf 4.4. Equations of motion}

 They take the form (see (\ref{H2}),(\ref{2.2}))
$$
\frac{\de A}{\de\mu}(z,\bar{z}) =0,~~
\ka \frac{\de \bA}{\de\mu}(z,\bar{z})=A(z,\bar{z}),
$$
\beq{3.2}
\frac{\de p_b}{\de\mu}=0,~~
\frac{\de g_b}{\de \mu}=0,
\eq

We can introduce the modified connection
\beq{4.2}
\bA'=\bA-\frac{1}{\ka}\mu A.
\eq
In its terms (\ref{2.2})  take the canonical form:
\beq{5.2}
\om=\int_{\Si}<\de A,\de \bA'>+\sum_{a=1}^n\de<p_a,g_a^{-1}\de g_a>.
\eq
and the equations of motion (\ref{3.2}) become trivial
\beq{6.2}
\frac{\de A}{\de\mu}(z,\bar{z}) =0,~~~
\frac{\de \bA'}{\de\mu  }(z,\bar{z})=0.
\eq
In other words, as we discussed in Sect.2, if we forget about
the complex structures of curves the bundle
become trivial ${\cal P}'\sim{\cal R}'\times{\cal N}'$.
\bigskip

 {\bf 4.5. Symmetries}

The form $\om$ (\ref{2.2}) (or (\ref{5.2})) is invariant with respect
to the action of
the group ${\cal G}_0$ of  diffeomorphisms of $\Si_{g,n}$,
which are trivial in vicinities ${\cal U}_a$ of marked points:
\beq{7.2}
{\cal G}_0=\{z\rar N(z,\bar{z}),\bar{z}\rar \bar{N}(z,\bar{z}),),~
 N(z,\bar{z})=z+o(|z-x_a|),~z\in {\cal U}_a\}.
\eq
In particular, its action on the Beltrami differentials takes the
 form of the M\"{o}bius transform
\beq{7a.2}
\mu\rar\frac
{\frac{\p z}{\p \bar{N}}+\mu\frac{\p \bar{z}}{\p \bar{N}}}
{\frac{\p z}{\p N}+\mu\frac{\p \bar{z}}{\p N}}
\eq

Another infinite gauge symmetry of  the form (\ref{2.2})
(or (\ref{5.2}))  is the
group
$${\cal G}_1=\{f(z,\bar{z})\in C^\infty (\Si_{g},G)\},$$
that acts on the dynamical fields as
$$
A+\ka\p\rar f(A+\ka\p)f^{-1},~~
\bar{A}+\bp+\mu\p\rar f(\bar{A}+\bp+\mu\p)f^{-1},
$$
\beq{8.2}
(\bA'+\bp\rar f(\bar{A}'+\bp)f^{-1}),
\eq
$$
p_a\rar f_ap_af^{-1}_a,
~~g_a\rar g_af^{-1}_a,~~(f_a=\lim_{z\rar x_a}f(z,\bz)),
~~
\mu\rar\mu.
$$
In other words, the gauge action of ${\cal G}_1$
 does not touch the base  ${\cal N}'$ and transforms only the
fibers ${\cal R}'$.
The whole gauge group is the semidirect product
\beq{9.2}
{\cal G}_1\oslash{\cal G}_0.
\eq

There is an additional finite-dimensional symmetry group ${\cal G}_2$
   which commutes with (\ref{9.2}). It acts
only on the singular curves in the fibers at the marked points.
It is remnant of ${\cal G}_1$ on the desingular curves
 (see Sect.2).
$${\cal G}_2=\otimes_{a=1}^nG_a,$$
\beq{9a.2}
p_a\rar p_a,~~g_a\rar h_ag_a,~~h_a\in G_a,~(a=1,\ldots,n).
\eq
This action commutes with (\ref{9.2}).
\bigskip

{\bf 4.6. Symplectic reduction with respect to} ${\cal G}_1$

The infinitesimal action of ${\cal G}_1$ (\ref{8.2}) generates
the vector field
\mbox{$\ep_1\in {\cal A}^{(0)}(\Si_{g,n},{\rm Lie}(G))$}
$$j_{\ep_1}A=-\ka\p\ep_1 +[\ep_1,A],$$
$$j_{\ep_1}\bA'=-\bp\ep_1+[\ep_1, \bA'],$$
$$ j_{\ep_1}p_a=[\ep_1(z_a,\bz_a),p_a],$$
$$j_{\ep_1}g_a=-g_a\ep_1(z,\bz_a),$$
$$j_{\ep_1}\mu=0.$$
 Let
$$
F_{A,\bA'}=\bp A-\ka\p\bA'+[\bA,A]=(\bp +\p\mu)A-\ka\p\bA+[\bA.A].
$$
Note that the action of the operator $(\bp+\p\mu)$ on the connection
$A$ is correctly defined.

The corresponding moment map
$$
{\cal J}_1:{\cal P}\rar {\rm Lie}^*({\cal G}_1).
$$
takes the form
$$
{\cal J}_1=-F_{A,\bA'}(z,\bz)+2\pi i\sum_{a=1}^n\de^2(x^0_a)p_a.
$$
Let ${\cal J}_1=0.$ In other words, the moment constraints
equation takes the form
\beq{10.2}
F_{A,\bA'}(z,\bz)=2\pi i\sum_{a=1}^n\de^2(x^0_a)p_a.
\eq
It means that we deal with the flat connection
everywhere on $\Si_{g,n}$ except
the marked points. The holonomies of $(A,\bA)$ around the marked
points are conjugated to  $\exp 2\pi ip_a$.

Let $(L,{\bar L})$ be the gauge transformed connections
\beq{11.2}
\bA=f\bar{L}f^{-1}+f(\bp+\mu\p)f^{-1},
\eq
\beq{12.2}
A=fLf^{-1}+\ka f\p f^{-1},
\eq
Then (\ref{10.2}) takes the form
\beq{12a.2}
(\bp+\p\mu)L-\ka\p {\bar L}+[\bar{L},L]=
2\pi i\sum_{a=1}^n\de^2(x^0_a)p_a.
\eq

The gauge fixing allows to choose $\bA$ in a such way
that $\p\bar L=0$. In fact,  the antiholomorphity of
$f^{-1}(\bp +\mu\p)f+f^{-1}\bar{A}f$  amounts  to the classical
equations of motion for
the Wess-Zumino-Witten functional $S_{WZW}(f,\bar{A})$
in the external field ${\bar A}$, which does  has extremal points.
Then instead of (\ref{12a.2}) we have
 \beq{14.2}
(\bp+\p\mu)L+[\bar{L},L]=\sum_{a=1}^n\de^2(x_a)p_a.
\eq
We can rewrite it as
\beq{13.2}
\p_{\bar w}L+[\bar{L},L]=
2\pi i\sum_{a=1}^n\de^2(x_a)p_a.
\eq
Anyway, by choosing $\bar{L}$ we fix somehow the gauge
in generic case. The last form of the moment constraint
(\ref{13.2}) coincides with  that for the Hitchin systems \cite{Ne},
which allows to apply the known solutions.

\bigskip
{\bf 4.7. Symplectic reduction with respect to} ${\cal G}_2$

The gauge transforms $h_a\in G_a$ in the points $x_a$  acts on
$T^*G_a$. In the case of punctures it allows to fix $p_a$
on some coadjoint
orbit $p_a=g_ap_a^{(0)}g_a^{-1}$ and obtain the symplectic quotient
${\cal O}_a=T^*G_a//G_a$. In fact, the moment corresponding
to this action is
$$\mu_a=g_ap_ag_a^{-1}\in{\rm Lie}^*(G).$$
Let $\mu_a=J_a$ be some fixed point in ${\rm Lie}^*(G).$
Then the gauge fixing
allows to choose $g_a$ up to the stabilizator of $J_a$ and
$$p_a=g_a^{-1}J_ag_a\in {\cal O}_a.$$
Thus, in (\ref{12a.2}) or (\ref{14.2}) $p_a$ are
elements of ${\cal O}_a$. The symplectic form on ${\cal O}_a$ keeps
 the same form as on $T^*G_a$.

Consider the double points case (the nodal singularities).
 Let $x_1$ and $x_2$ be preimages of the
nodal point $y$ under the normalization map. Then the symplectic form
on the normalization of the singular curve
$$
\om=\de<p_1,g^{-1}_1\de g_1>+\de<p_2,g^{-1}_2\de g_2>
$$
generates the moment
$$
\mu=g_1p_1g_1^{-1}+g_2p_2g_2^{-1}.
$$
Put $\mu=0$. Then
$$
p_1=-\ti{g}_2p_2\ti{g}_2^{-1},~~\ti{g}_2=g_1^{-1}g_2.
$$
The pair $(p_2,\ti{g}_2)$ is an arbitrary element of $T^*G$.
Therefore, in this case \\$T^*G_1\oplus T^*G_2//G=T^*G$.

In what follows we will concentrate on the case of curves
with punctures (without double points).  Let ${\cal I}_{g,n}$
be the equivalence classes of the connections
$(A,\bA)$ with respect to the gauge action (\ref{11.2}),(\ref{12.2}) -
 the moduli space of stable flat $G$ bundles
over $\Si_{g,n}$.  It is a smooth
finite dimensional space.
  If we fix the conjugacy classes of holonomies
$(L,\bar L)$ around marked points
${\cal I}_{g,n}$ becomes a symplectic manifold.
It is extended here by the coadjoint
 orbits ${\cal O}_a$  in the marked points
$x_a,(a=1,\ldots,n)$  in the consistent way (see (\ref{12a.2})).
Fixing the gauge we come to the symplectic quotient
$${\cal R}\subset{\cal I}_{g,n}\times\prod_{a=1}^n{\cal O}_a,$$
$${\cal R}={\cal R}'//({\cal G}_1\oplus {\cal G}_2)
={\cal J}^{-1}_1(0)/({\cal G}_1\oplus {\cal G}_2).$$
It has  dimension
$$
\dim({\cal R})=
\left\{
\begin{array}{ll}
\dim(\sum_{a=1}^n{\cal O}_a//G),&~g=0,\\
2{\rm rank} G+\dim(\sum_{a=1}^n{\cal O}_a//H),&~g=1\\
(2g-2)\dim G+\dim(\sum_{a=1}^n{\cal O}_a)&~g\geq 2,\\
\end{array}
\right.
$$
where $H$ is  the Cartan subgroup $H\subset G$  and ${\cal O}_a//G$
and ${\cal O}_a//H$ are the symplectic quotients of the symplectic spaces
${\cal O}_a$ under the actions
of the automorphisms of the bundles in the rational and the elliptic cases
correspondingly.
 The connections $(L,\bar{L})$ in addition
to ${\bf p}=(p_1,\ldots,p_n)$ depend
 on  a finite even number of free parameters $2r$
$$
({\bf v},{\bf u}),~{\bf v}=(v_1,\ldots,v_{r}),~{\bf u}=(u_1,\ldots,u_{r}).
$$
$$
r=
\left\{
\begin{array}{ll}
0&~g=0,\\
{\rm rank} G,&~g=1,\\
(g-1)\dim G,&~g\geq 2.\\
\end{array}
\right.
$$
${\cal R}$ is a symplectic manifold with the non degenerate symplectic form
which is the reduction of (\ref{1.2})
\beq{15.2}
\om_0=\int_{\Si}<\de L,\de \bar{L}>+
2\pi i\sum_{a=1}^n\de<p_a,g_a\de g_a^{-1}>=
\eq

On this stage we come to the bundle ${\cal P}''$
with the finite-dimensional
fibers
 ${\cal R}$ over the infinite-dimensional
base ${\cal N}'$  with the symplectic form
\beq{16.2}
\om=\int_{\Si}<\de L,\de\bar{L}>+
2\pi i\sum_{a=1}^n\de<p_a,g_a^{-1}\de g_a>-\ka\int_{\Si}<L,\de L>\de\mu.
\eq
\bigskip

{\bf 4.8. Factorization with respect to the diffeomorphisms} ${\cal G}_0$

We can utilize invariance of $\om$ with respect to ${\cal G}_0$
and reduce ${\cal N}'$ to the finite-dimensional
space ${\cal N}$, which is isomorphic to the moduli space
${\cal M}_{g,n}$.
 Let $\ep_0$ be a vector field generated by the diffeomorphisms
(\ref{7.2}).
 Consider the action of the Lie derivative
${\cal L}_{\ep_0}$ on ${\cal A}=(A,\bA)$
$${\cal L}_{\ep_0}{\cal A}=dj_{\ep_0}{\cal A}+j_{\ep_0}d{\cal A}=
d_{\cal A}(j_{\ep_0}{\cal A})+j_{\ep_0}F_{\cal A},~~
(F_{\cal A}=d{\cal A}+\frac{1}{2}[{\cal A},{\cal A}]).$$
Thus for the flat connections $F_{\cal A}=0$ the action
of diffeomorphisms ${\cal G}_0$ on the connection fields
is generated by the gauge transforms $j_{\ep_0}{\cal A}\in{\cal G}_1$.
But we already have performed
the symplectic reduction with respect to ${\cal G}_1$. Therefore,
$j_{\ep_0}$ belongs to the kernel of $\om$ (\ref{16.2})
and we can push it
down on the factor space ${\cal P}''/{\cal G}_0$.
 Since ${\cal G}_0$ acts only  on
${\cal N}'$, it can be done by  fixing
the dependence of $\mu$ on the coordinates
 in the  Teichm\"{u}ller space
 ${\cal T}_{g,n}$.
According to (\ref{2b.2}) represent $\mu$ as
\beq{17a.2}
\mu=\sum_{s=1}^{l}\mu_s,~~\mu_s=t_s\mu_s^0,~l=\dim {\cal T}_{g,n},~
\mu_s^0=\bp n_s.
\eq
The Beltrami differential (\ref{17a.2}) defines the
tangent vector
$$
{\bf t}=(t_1,\ldots,t_l),
$$
to the Teichm\"{u}ller  space  ${\cal T}_{g,n}$ at the
fixed point of ${\cal T}_{g,n}$.

We specify the dependence of $\mu$ on the positions of
the marked points in the following  way.
Let ${\cal U}'_a\supset{\cal U}_a$ be two vicinities
 of the marked point $x_a$
such that ${\cal U}'_a\cap{\cal U}'_b=\emptyset$ for $a\neq b$.
Let $\chi_a(z,\bz)$ be a smooth function
\beq{cf}
\chi_a(z,\bz)=\left\{
\begin{array}{cl}
1,&\mbox{$z\in{\cal U}_a$ }\\
0,&\mbox{$z\in\Si_g\setminus {\cal U}'_a.$}
\end{array}
\right.
\eq
Introduce times related to the positions of the
marked points $t_a=x_a-x_a^0$. Then
\beq{17.2}
\mu^0_a=\bp n_a(z,\bz),~~n_a(z,\bz)=(1+c_a(z-x_a))\chi_a(z,\bz)
\eq
In other words $n_a(z,\bz)$ defines
a local vector field deforming the complex
coordinates only in ${\cal U}'_a$
$$w=z-t_an_a(z,\bz).$$

The action of ${\cal G}_0$ on the phase space ${\cal P}''$
reduces the infinite-dimensional component ${\cal N}'$ to
${\cal T}_{g,n}$.
After the reduction we come to the bundle with base ${\cal T}_{g,n}$.
Substituting
\beq{18.2}
\de\mu=\sum_{s=1}^l\mu_s^0\de t_s,~~(\p_s=\p_{t_s}).
\eq
in(\ref{16.2}) we obtain
\beq{19.2}
\om=\om_0({\bf v},{\bf u},{\bf p},{\bf t})-
\frac{1}{\ka}\sum_{s=1}^{l}
\de H_s({\bf v},{\bf u},{\bf p},{\bf t})\de t_s,
\eq
where $\om_0$ is defined by (\ref{15.2}), and $H_s$ are the Hamiltonians
\beq{20.2}
H_s=\int_\Si <L,L>\p_s\mu
\eq

In fact, we still have a remnant discrete  symmetry, since $\om$
 is invariant under the
mapping class group  $\pi_0({\cal G}_0)$. Eventually, we come to
the moduli space
 ${\cal M}_{g,n}={\cal T}_{g,n}/\pi_0({\cal G}_0).$

Summarizing, we have defined the extended phase space
(the bundle ${\cal P}$) as the result of the symplectic reduction
with respect to the ${\cal G}_1\oplus{\cal G}_2$ action and subsequent
 factorization under the ${\cal G}_0$ action.
We can write symbolically
$$
{\cal P}=({\cal P}''//{\cal G}_1\oplus{\cal G}_2)/{\cal G}_0.
$$
It is endowed with the symplectic form (\ref{19.2}).
\bigskip

{\bf 4.9. The hierarchies of the isomonodromic deformations (HID)}

The equations of motion can be extracted from the symplectic form
(\ref{19.2}), as it was described in Section 3 (see (\ref{H2})).
They will be referred as {\sl the hierarchies of the isomonodromic
deformations} (HID). This notion will be justify later.
In terms of the local coordinates
(\ref{H2}) takes the form
\beq{21.2}
\ka\p_s{\bf v}=\{H_s,{\bf v}\}_{\om_0},~~
\ka\p_s{\bf u}=\{H_s,{\bf u}\}_{\om_0},~~
\ka\p_s{\bf p}=\{H_s,{\bf p}\}_{\om_0}~~
\eq
The Poisson bracket $\{\cdot,\cdot\}_{\om_0}$
is the inverse  tensor to $\om_0$.
We also has the {\sl Whitham hierarchy} (\ref{H3}) accompanying
(\ref{21.2}).
It follows from (\ref{15.2}) that the Hamiltonians
$H_s$ (\ref{20.2}) commute.
$$
\{H_r,H_s\}_{\om_0}=0.
$$
Therefore,
\beq{22.2}
\p_s H_r-\p_r H_s=0,
\eq
and there exists the one form on ${\cal M}_{g,n}$ defining {\sl the tau
function} of the hierarchy of isomonodromic deformations
\beq{22a.2}
\de \log\tau=-\f1{\ka}\sum H_sdt_s.
\eq

The following three statements are valid for the hierarchy of
isomonodromic deformations (\ref{21.2}):\\
\begin{predl}
The flatness condition (\ref{12a.2}) and HID (\ref{21.2}) are equivalent
to the consistent system of linear equations
\beq{23.2}
(\ka\p+L)\Psi=0,
\eq
\beq{24.2}
(\ka\p_s+M_s)\Psi=0,~~(s=1,\ldots,l=\dim{\cal M}_{g,n})
\eq
\beq{25a.2}
(\bp+\mu\p+{\bar L})\Psi=0,
\eq
 where $M_s$ is a solution to the linear equation
\beq{25.2}
\p_{\bar w}M_s-[M_s,{\bar L}]=\ka\p_s\bar{L}-L\mu^0.
\eq
\end{predl}

\begin{predl}
The linear conditions (\ref{24.2})
 provide the isomonodromic deformations of the linear system (\ref{23.2}),
(\ref{25a.2})
 with respect to change the "times" on ${\cal M}_{g,n}$.
\end{predl}
Therefore, HID (\ref{21.2}) are the monodromy preserving
conditions for the linear system (\ref{23.2}),(\ref{25a.2}).\\

The presence of derivative with
respect to the spectral parameter $w\in\Si_{g,n}$ in the linear equation
(\ref{23.2}) is a
distinguish feature of the isomonodromy preserving equations.
It plagues  the
application of the inverse scattering method to this type of systems.
The later means the solving the Riemann-Hilbert problem which
amounts the reconstructing the pair $(L,\bar L)$ from the
monodromy data. In \cite{IN} this technique was applied to calculate
asymptotics of solutions for bundles over rational curves.
Nevertheless, in general case
 we have in some sense the explicit form of solutions:
\begin{predl}[The projection method.]
The solution of the Cauchy problem of (\ref{21.2})
 for the initial data  ${\bf v}^0,{\bf u}^0,{\bf p}^0$  at the time
 ${\bf t}={\bf t}^0$
 is defined in terms of the elements $L^0,{\bar L}^0$  as the gauge
transform
\beq{26a.2}
{\bar L}({\bf t})=f^{-1}( L^0(\mu({\bf t})-
\mu({\bf t}^0))+{\bar L}^0)f+
f^{-1}(\bp+\mu({\bf t})\p)f,
\eq
\beq{26.2}
L({\bf t})=f^{-1}(\p+L^0)f,~~{\bf p}({\bf t})=f^{-1}({\bf p}^0)f,
\eq
 where $f=f(z,\bz)$  is a smooth  $G$-valued functions on $\Si_{g,n}$
fixing the gauge.
\end{predl}
It means that solutions of HID are gauge transformations of free motion
in the upstairs system.
Equations (\ref{26a.2}), (\ref{26.2}) look like the dressing
transform of the free motion. To find solutions one should know the gauge
transform $f$ from the upstairs system to a fixed gauge. For example,
in case of genus zero we consider the holomorphic solutions $\Psi$.
Therefore, $f$ should kill the ${\bar L}$ operator in (\ref{26a.2})
(see below).

\bigskip

{\sl Proofs.}\\
To prove first statement represent $A$ as (\ref{12.2}).
 The first equation
in (\ref{3.2}) $\p_s A=0$ means that
\beq{27.2}
\ka\p_sL-\ka\p M_s+[M_s,L]=0.
\eq
Then (\ref{27.2}) is the consistency condition
for the linear system (\ref{23.2}),
(\ref{24.2}).
 On this stage $M_s$ is defined as $M_s=\ka f^{-1}\p_sf$.
To find the
linear equation (\ref{25.2}), defining $M_s$, it is necessary to
substitute the gauge transformed
form  of $\bA$ (\ref{11.2}) together with (\ref{12.2})
in the equation of motion
 in the form $\ka\p_s\bA=A\p_s\mu$. This equality is the same as
$$
\ka(\p_s\mu\p+\p_s{\bar L})-\p_{\bar w}M_s+[M_s,\bar{L}]
=\p_s\mu(\ka\p+L).
$$
The later equation coincides with (\ref{25.2}). Eventually,
$$
\begin{array}{ccc}
{\rm the~ compatibility ~
(\ref{23.2}) ,(\ref{24.2})}&\longleftrightarrow &
 {\rm the~ Lax ~equations ~ (\ref{27.2})},\\
{\rm the~ compatibility ~
(\ref{23.2}),(\ref{25a.2})}&\longleftrightarrow&
{\rm  the ~flatness} (\ref{12a.2}),\\
{\rm the~ compatibility ~ (\ref{24.2}) ,(\ref{25a.2})}
&\longleftrightarrow&{\rm(\ref{25.2})}.
\end{array}
$$
This concludes the first statement.

To prove the second statement  note that (\ref{23.2}),(\ref{25a.2})
are equivalent to
$$
(\ka\p+A)\Psi^f=0,
$$
$$
(\bp+\mu\p+\bA)\Psi^f=0, ~~(\Psi^f=f^{-1}\Psi).
$$
Due to the equations of motion
\mbox{$(\p_sA=0,~\p_s\bA=\f1{\ka}A\p_s\mu)$} the monodromies of this
system are independent on moduli, i.e. $\p_s\Psi^f=0$. The monodromies
of the reduced system
(\ref{23.2}),(\ref{25a.2}) are conjugate
to the monodromies of the later one.
Thus, we come to the second statement.

To derive the expressions for dynamical variables
in the projection method we
lift the initial data $L^0,\bar{L}^0,p_1^0,\ldots,p_n^0$
from the reduced phase space ${\cal R}$ in the point $t^0$
to ${\cal R'}$ by the trivial gauge transform.
Due to the equations of motion (\ref{3.2}), the evolution in
${\cal R}$ is trivial
$$A(t)=L^0,~~\bA=\ka L^0(\mu(t)-\mu(t^0))+(\bar{L}^0)$$
and can be push back on the reduced phase
space ${\cal R}$ by the gauge transform to the fixed gauge.
This procedure is
reflected in projection method formulae (\ref{26a.2}),(\ref{26.2}).

\section{Remarks about the Hitchin systems and the KZB equations}
\setcounter{equation}{0}

{\bf 5.1. Scaling limit.}

Consider our system in the limit $\ka\rar 0$.
We will prove that
in this limit we come to
the Hitchin systems, which are living on the cotangent
bundles to the
moduli space of holomorphic $G$-bundles over $\Si_{g,n}$.
The value $\ka=0$ is called critical and looks singular
(see (\ref{2.2}),
(\ref{19.2})).
To get around we rescale the times
\beq{1.3}
{\bf t}=\bf T +\ka{\bf t}^H,
\eq
where  ${\bf t}^H$ are the fast (Hitchin) times and $\bf T$ are the
slow times. Therefore,
$$
\de\mu({\bf t})=\ka\sum_s\mu_s^0\de t_s^H,~~
(\mu_s^0=\bp n_s).
$$
After this rescaling the forms  (\ref{2.2}),(\ref{19.2})
 become regular.
The rescaling procedure means that we blow up a vicinity
 of the fixed point
(\ref{1.3}) in ${\cal M}_{g,n}$ and the whole dynamic of the Hitchin
systems is developed in this vicinity
\footnote{We are grateful to A.Losev for elucidating this point.}.
For example, we have instead (\ref{1.2}) and (\ref{2.2})
\beq{2.3}
\om=\int_{\Si}<\de A,\de\bar{A}>+
2\pi i\sum_{a=1}^n\de<p_a,g_a^{-1}\de g_a>-
\sum_s\int_{\Si}<\de A, A>\p_s\mu^0\de t^H.
\eq
If $\ka=0$ the connection $A$ behaves  as the one-form
$A\in \Om^{(1,0)}(\Si_{g,n},{\rm Lie}(G))$ (see (\ref{8.2})).
It is the so called
the Higges field in terms of \cite{H1}. An important point is that
the Hamiltonians now become the times independent.
The form (\ref{2.3}) is
the starting point in the derivation
of the Hitchin systems via the symplectic
reduction \cite {H1,Ne}. Essentially, it is the  same  procedure
as described above. Namely, we obtain
the same moment constraint (\ref{14.2})
and the same gauge fixing (\ref{11.2}).
But now we are sitting in a fixed point $\mu({\bf t}^0)$
of the moduli space ${\cal M}_{g,n}$
and don't need the factorization under
the action of the diffeomorphisms and,
thereby, do not worry about
 the modular properties of solutions $(L,\bar {L})$ of the
moment constraint (\ref{14.2}).
This only difference between the solutions
$(L,\bar{L})$ and the quadratic Hamiltonians $H_s$ in
the Hitchin systems and in HID. The symplectic reduction allows to
identify the phase space ${\cal R}$ with the cotangent bundle to
the moduli of holomorphic stable bundles.

Propositions 4.1 and 4.3 are valid
for the Hitchin systems in a slightly modified form.
\begin{predl}
 There exists the consistent system of
linear equations
\beq{3.3}
(\la+L)\Psi=0,~~\la\in\bf{C}
\eq
\beq{4.3}
(\p_s+M_s)\Psi=0,~~
\p_s=\frac{\p}{\p t^H_s},~(s=1,\ldots,l=\dim{\cal M}_{g,n})
\eq
\beq{5.3}
(\bp+{\bar L})\Psi=0, ~\bp=\p_{\bz},
\eq
 where $M_s$ is a solution to the linear equation
\beq{6.3}
\bp M_s-[M_s,{\bar L}]=\p_s\bar{L}-L\mu^0.
\eq
\end{predl}

 Here we have
$$
\begin{array}{ccc}
{\rm The~ compatibility ~
(\ref{3.3}) ,(\ref{4.3})}&\longleftrightarrow &
 {\rm the ~Lax ~equation}~
\p_{T_s}L=[L,M_s],\\
{\rm the~ compatibility ~
(\ref{3.3}),(\ref{5.3})}&\longleftrightarrow &
{\rm  the ~moment~equation~(\ref{14.2})},\\
{\rm the~ compatibility ~ (\ref{4.3}) ,(\ref{5.3})}
&\longleftrightarrow&{\rm(\ref{6.3})}.
\end{array}
$$

To derive these equations from the general case (\ref{23.2}),
(\ref{24.2}), and (\ref{25a.2}) we use the WKB approximation
$$
\Psi=\Phi\exp\frac{\cal S}{\ka},
$$
where $\Phi$ is a group valued function and
$${\cal S}={\cal S}({\bf t},w)={\cal S}^0({\bf T},w^0)+
\ka{\cal S}^1,~~w^0=z-\sum_sT_sn_s(z,\bz),
$$
\beq{S1}
{\cal S}^1=\sum_st_s^H(\frac{\p}{\p_{T_s}}-n_s(z,\bz)\p){\cal S}^0.
\eq
 In the first order in $\ka$ the equations
 (\ref{23.2}),(\ref{24.2}),(\ref{25a.2}) just gives
 (\ref{3.3}),(\ref{4.3}),(\ref{5.3}) if
 \beq{SW}
 \p{\cal S}^0=\la,
\eq
 $$
\frac{\p}{\p\bar w_0}{\cal S}^0=0,~~
\frac{\p}{\p t^H_s}{\cal S}^0=0.
$$
Therefore, as soon as ${\cal S}^0$ satisfies these equation,
$\Psi$ provides solutions to the linearization of Hitchin systems.
The detaile analyses of the perturbation in the rational
case was undertaken in \cite{Ta}.

 Equation (\ref{3.3}) allows to introduce the fixed spectral curve
 $$
 {\cal C}:~\det(\la\cdot Id+L)=0,~~{\cal C}\in {\bf P}(T^*\Si\oplus 1)
 $$
 where $\la$ is a coordinate in the cotangent space. The Hitchin phase
 space ${\cal R}$ has the "spectral" description. It is the bundle
 $$
 \pi:{\cal R}\rar {\cal M}_{C}
 $$
over the moduli ${\cal M}_C$ of spectral curves
with abelian varieties as generic fibers.
 The map $\pi$ acts from the pair $(A,\bA)\sim (L,\bar L)$, ($L$ now
 is the Higgs field
 $L\in \Om^{(1,0)}(\Si_{g,n},{\rm Lie}^*(G))$)
 to the set of coefficients of
 the characteristic polynomial $\det(\la\cdot Id+L)$.
 The one-form $\te=\la dw^0$ being integrated over
 corresponding cycles in ${\cal C}$ gives rise to the action variables.
 The angle variables can be also extracted from the spectral curve.
 All together defines the symplectic structure
 on the Hitchin phase space
 in the spectral picture. This original
 Hitchin construction is working in the singular case as well
 \cite{Ne}.
 The symplectic structure of this type connected
 with hyperelliptic curves
 was introduced originally in the soliton theory by Novikov and Veselov
  \cite{VN}.
 In terms of 4d gauge theories $\te$ is the Seiberg-Witten
 differential. It follows from (\ref{SW}) that $\te=d{\cal S}^0$ and
 along with (\ref{S1}) it defines the first order approximation to the
 solutions of the
 linear form (\ref{23.2}),(\ref{24.2}),(\ref{25a.2}) of HID.

 It is possible to define the dynamic of the spectral curve beyond the
 crtical level as it was done for the Painlev\'e equations in \cite{V}.
 It follows from the Lax representation (\ref{27.2}) that
 $$
\p_s \det(\la\cdot Id+L)=\tr\p_wM_s(\la\cdot Id+L)^{-1}.
 $$
 Note, that it defines the motion of ${\cal C}$ only within
 the subset
 ${\cal M}_{g,n}\subset {\cal M}_C$.

When $L$ and thereby $M$ can be find explicitly the simplified
form of (\ref{14.2})
allows to apply the inverse scattering method
to find  solutions
of the Hitchin hierarchy as it was done for $\SL$ holomorphic
bundles over $\Si_{1,1}$ \cite{Kr2}, corresponding to
the elliptic Calogero system. We present the alternative way
to describe the solutions:\\
\begin{predl}[The projection method.]
$$
\bar{L}(t_s)=f^{-1}(L^0(t_s-t_s^0)\p_s\mu^0+
\bar{L}^0)f+f^{-1}\bp f,
$$
$$
L(t)=f^{-1}L^0f,~~p_a(t)=f^{-1}(p_a^0)f
$$
\end{predl}
The degenerate version of these expressions was known
for a long time \cite{OP}.
\bigskip
\noindent

{\bf 5.2. About KZB}

The Hitchin systems are the classical limit of the KZB equations
on the critical level \cite{Ne,I2}.
The later has the form of the Schr\"{o}dinger
equations, which is the result of geometric quantization of the
moduli of flat $G$
bundles \cite{ADPW,H2}. The conformal blocks of the
WZW theory on $\Si_{g,n}$
with vertex operators in marked points
are the ground state wave functions
$$\hat {H}_sF=0,~~(s=1,\ldots,l).$$
The classical limit means that one replaces operators
on their symbols
and generators of
finite-dimensional representations in the vertex operators
 by the corresponding
elements of coadjoint orbits.

Generically, for the quantum level $\ka^{quant}\neq 0$
the KZB equations can be written in the form of the
nonstationar  Schr\"{o}dinger equations \cite{H2,I2}
$$
(\ka^{quant}\p_s+\hat {H}_s)F=0.
$$
To pass to the classical limit in this equation
we replace the conformal block by its quasiclassical expression
$$F=\exp \frac{iS}{\hbar},$$
where $S$ is the classical action
($S=\log\tau$ (\ref{22a.2})) and renormalize
$$\ka=\frac{\ka^{quant}}{\hbar}.$$
The classical limit $\hbar\rar 0,~\ka^{quant}\rar 0$ leads the
Hamilton-Jacobi equations for S, which are equivalent to the
HID (\ref{21.2}).

Summarizing, we arrange these
quantum and classical systems in the commutative diagram.
The vertical
arrows denote to the classical limit and mean the simultaneous
rescaling of the quantum level, while  the limit
 $\ka^{quant}\rar 0$
$(\ka\rar 0)$
 on the
horizontal arrows includes also the rescaling
the moduli of complex structures.
The examples in the bottom of the diagram will
be considered in next sections.
$$
\def\normalbaselines{\baselineskip20pt
       \lineskip3pt    \lineskiplimit3pt}
\def\mapright#1{\smash{
        \mathop{\longrightarrow}\limits^{#1}}}
\def\mapdown#1{\Big\downarrow\rlap
       {$\vcenter{\hbox{$\scriptstyle#1$}}$}}
\begin{array}{ccc}
\left\{
\begin{array}{c}
\mbox{KZB eqs.},~(\ka,{\cal M}_{g,n},G)\\
(\ka^{quant}\p_{t_a}+\hat{H}_a)F=0,\\
(a=1,\ldots,\dim{\cal M}_{g,n})
\end{array}
\right \}
  &\mapright{\ka^{quant}\rar0} &
\left\{
\begin{array}{c}
\mbox{KZB eqs. on the critical level},~\\
({\cal M}_{g,n},G),~(\hat{H}_a)F=0,\\
(a=1,\ldots,\dim{\cal M}_{g,n})
\end{array}
\right\}
\\
\mapdown{\hbar\rar 0}&    &\mapdown{\hbar\rar 0} \\
\end{array}
$$
$$
\def\normalbaselines{\baselineskip20pt
       \lineskip3pt    \lineskiplimit3pt}
\def\mapright#1{\smash{
        \mathop{\longrightarrow}\limits^{#1}}}
\def\mapdown#1{\Big\downarrow\rlap
       {$\vcenter{\hbox{$\scriptstyle#1$}}$}}
\begin{array}{ccc}
\left\{
\begin{array}{c}
\mbox{Hierarchies of Isomonodromic} ~\\
\mbox{deformations on}~{\cal M}_{g,n}
\end{array}
\right \}
&\mapright{\ka\rar 0} &
\left\{
\begin{array}{c}
\mbox{Hitchin systems} ~\\
       \\
 \end{array}
\right \}
\\
 & & \\
  & \mbox{\sl EXAMPLES}& \\
\left\{
\begin{array}{c}
\mbox{Schlesinger eqs.} \\
\mbox{Painlev\'{e} type eqs.} \\
\mbox{Elliptic Schlesinger eqs.}
\end{array}
\right\}
  &  \mapright{\ka\rar 0} &
\left\{
\begin{array}{c}
\mbox{Classical Gaudin eqs.}\\\
\mbox{Calogero eqs.}\\
\mbox{Elliptic Gaudin eqs.}
\end{array}
\right\}
\end{array}
$$

\section{Genus zero -  Schlesinger equations}
\setcounter{equation}{0}
{\bf 6.1. Derivation of equations}

Consider ${\bf C}P^1$ with $n$ punctures
$(x_1,\ldots,x_n|x_a\neq x_b)$.
The Beltrami
differential $\mu$ is related only to the positions of marked points.
Then from (\ref{17.2})
\beq{2.0}
\de\mu_a=\bp (1+c_a\bp(z-x_a)\chi_a(z,\bz))\de t_a,~~(\de t_a=\de x_a).
\eq

On ${\bf C}P^1$  the gauge
transform (\ref{11.2}) allows to choose $\bA$ to be identically zero.
After  the gauge fixing
\beq{GF}
\bA=f\p_{\bar{w}}f^{-1},
\eq
$$A=fLf^{-1}+\ka f\p_wf^{-1},$$
the moment equation takes the form
$$
\p_{\bar w} L=2\pi i\sum_{a=1}^n\de^2(x_a)p_a.
$$
 It allows to find $L$
\beq{L}
L=\sum_{a=1}^n\frac{p_a}{w-x_a}.
\eq
Then we have from (\ref{2.0})
$$
\frac{1}{2}\de\int_{{\bf C}P^1}<L,L>\de\mu=
$$
$$
=\frac{1}{2}\sum_{b,a}\de\int_{{\bf C}P^1}
\frac{<p_a,p_b>}{(w-x_b)(w-x_a)}\sum_c\bp(1+c_c(w-x_c))\chi_c
\de x_c=
\sum_a(\de H_{a,1}+\de H_{a,0})\de x_a,
$$
where
\beq{3.5}
H_{a,1}=\sum_{b\neq a}\frac{<p_a,p_b>}{x_a-x_b},
\eq
and
$$H_{2,a}=c_a<p_a,p_a>.$$
$H_{1,a}$ are
precisely the Schlesinger's Hamiltonians.
On the symplectic quotient $\om$ (\ref{16.2})
takes the form
$$\om=\sum_{a=1}^n\de <p_a,g_a^{-1}\de g_a>-
\frac{1}{\ka}\sum_{b=1}^n(\de H_{b,1}+\de H_{b,0})\de x_b.
$$
Note, that we still have a gauge freedom with respect
to the coordinate
independent $G$ action, since this action
does not change our gauge fixing
(\ref{GF}).
The corresponding moment constraint means that the sum of residues
of $L$ vanishes:
\beq{VP}
\sum_{a=1}^np_a=0.
\eq

While $H_{2,a}$ are the Casimirs and lead to the trivial equations,
the equation of motion for $H_{1,a}$ are the Schlesinger equations
$$
\ka\p_bp_a=\frac{[p_a,p_b]}{x_a-x_b},~~(a\neq b),
$$
$$
\ka\p_ap_a=-\sum_{b\neq a}\frac{[p_a,p_b]}{x_a-x_b}.
$$
As by product we obtain by this procedure the corresponding
linear problem
(\ref{23.2}),(\ref{24.2}) and (\ref{25a.2}) with $L=$ (\ref{L}),
 $\bar{L}=0$ and
$$M_{a,1}=-\frac{p_a}{w-x_a}$$
 as a solution to (\ref{25.2}).
The tau-function (\ref{22a.2}) for the Schlesinger equations
has the form \cite{JMU}
$$
\de\log\tau=-\sum_{c\neq b}<p_b,p_c>\de\log (x_c-x_b).
$$

\bigskip
{\bf 6.2. Solutions via the projection method}

We will find the dressing transform defining the evolution on
the coadjoint orbits
\beq{dress}
p_a(\bft)=f^{-1}(z,\bz,\bft)p_a^0f(z,\bz,\bft).
\eq
Recall that the times in the Schlesinger equations are $t_a=x_a-x_a^0$
and assume that $t_a^0=0$. We have from (\ref{L})
$$
L(\bft_0=0)=L^0=\sum_{a=1}^n\frac{p^0_a}{z-x^0_a}.
$$
It follows from (\ref{GF}) that $\bar {L}=0$. Then
the  projection method (\ref{26a.2}) gives in this case
$$
f^{-1}(z,\bz,\bft)L^0\mu(\bft)f(z,\bz,\bft)+
f^{-1}(z,\bz,\bft)(\bp+\mu(\bft)\p)f(z,\bz,\bft)=0.
$$
In other words, the gauge transform $f(z,\bz,\bft)$ defining
the evolution of solutions (the dressing transform)
can be found from the equation
\beq{LE}
[\bp+\sum_kt_k\bp \chi_k(z,\bz)(\p+L^0)]f(z,\bz,\bft)=0.
\eq
We seek for smooth solutions to this equation assuming that the times
$t_s=x_s-x_s^0$ are small. To this end consider the perturbative series
\beq{pert}
f(z,\bz,\bft)=id+\sum_kt_ka_k+\sum_{j\leq k}t_jt_ka_{jk}+
\sum_{i\leq j\leq k}t_it_jt_ka_{ijk}+\ldots,
\eq
where $a_k,a_{jk},a_{ijk},\ldots$ are smooth maps ${\bf C}P^1\rar$
Universal enveloping algebra$(G)$.
Then (\ref{LE}) leads to the system
$$
\begin{array}{l}
1)\bp a_k=-L^0\bp\chi_k\\
2)\bp a_{jk}=-(\p+L^0)(\bp\chi_ja_k+\bp\chi_ka_j)\\
3)\bp a_{ijk}=-(\p+L^0)\bp\chi_{[j}a_{jk]}\\
4)\ldots,\\
\end{array}
$$
where $_{[\cdots]}$ means the symmetrization.
All equations have the same structure - their solutions depends only
on the previous step.
Since $(0,1)$-forms on ${\bf C}P^1$ are exact the
equations can be integrated and solutions are found step by step.

 In the first order one has
\beq{fo}
a_k(z,\bz)=-\sum_{a=1}^n\frac{p^0_a}{z-x^0_a}\chi_k^a(z,\bz), ~~
(\chi_k^a(z,\bz)=\chi_k(z,\bz)-\chi_k(x_a,\bar{x}_a)).
\eq
Note that $a_k(z,\bz)$ is a nonsingular function on
${\bf C}P^1$ due to the
definition of $\chi_k^a(z,\bz)$.  Consider now the second order
approximation.  $$ \bp a_{jk}(z,\bz)=\sum_{a=1}^n\p(\frac{p^0_a}{z-x^0_a})
\bp\chi_{[j}(z,\bz)\chi_{k]}^a(z,\bz)+
\sum_{a=1}^n\frac{p^0_a}{z-x^0_a}\bp\chi_{[j}(z,\bz)\p\chi_{k]}^a(z,\bz)+
$$
$$
+\sum_{a=1}^n\frac{p^0_a}{z-x^0_a}\sum_{b=1}^n\frac{p^0_b}{z-x^0_b}
\bp\chi_{[j}(z,\bz)\chi_{k]}^b(z,\bz).
$$
The result of integration is
\beq{so}
a_{jk}(z,\bz)=\sum_{a=1}^np^0_a(\frac{\psi_j^a(z,\bz)\de_{jk}}{z-x^0_a}-
\frac{\chi_j^a(z,\bz)\chi_k^a(z,\bz)}{(z-x^0_a)^2})+
\eq
$$
\sum_{a=1}^n\frac{(p^0_a)^2}{(z-x^0_a)^2}\chi_j(z,\bz)\chi_k^a(z,\bz)
-\sum_{a\neq b}^n\frac{p^0_ap_b^0}{z-x^0_b}
(\chi_j^a(z,\bz)\chi_k^a(z,\bz)-\chi_j^a(x_b^0)\chi_k^a(x_b^0)).
$$
Here $\psi_j^a(z,\bz)$ is defined as the result of integration
\beq{psi}
\bp\psi_j^a(z,\bz)=2\chi_j(z,\bz)\chi_j^a(z,\bz),~~(\psi_j^a(x_a)=0).
\eq
It provides the absence of poles in $z=x_a$ in the first term in
the right hand side of (\ref{so}).
Due to the subtraction of $\chi_j^a(x_b^0)\chi_k^a(x_b^0)$ in
the last term we kill the pole of $a_{jk}(z,\bz)$ in $z=x_b$. Thus,
in second order approximation
we obtain the regular solution as well. It is defined almost
explicitly up to the integration (\ref{psi}).

Therefore, we have defined the dressing transformation
(\ref{pert}),(\ref{fo}),(\ref{so}) up to the third order
of the initial data (\ref{dress}).
The calculations of the higher order corrections reproduce
the same procedure as
on the second order and we can repeat them step by step.

\section{Genus one- elliptic Schlesinger, Painlev\'{e} VI...}
\setcounter{equation}{0}

The genus one case is still feasible to write down
the explicit formulae for
the Hamiltonians and the equations of motion.\\
{\bf 7.1. Deformations of elliptic curves}

In addition to the moduli coming from the positions
of the marked points there
is the elliptic module $\tau,~Im\tau>0$ of curves $\Si_{1,n}$.
As in (\ref{17a.2}),(\ref{17.2}) we take
the Beltrami differential in the form
$$
\mu=\sum_{a=1}^n\mu_a+\mu_\tau,~~(\mu_a=t_a\bp n_a)
$$
where $n_a(z,\bz)$ is the same as in (\ref{17.2}) and
\beq{2.6}
n_\tau=(\bz-z)(1-\sum_{a=1}^n\chi_a(z,\bz)).
\eq
We replace
$$
t_\tau\rar \frac{t_\tau}{\rho},~~t_\tau=\tau-\tau_0,~~
\rho=\tau_0-\bar{\tau}_0.
$$
Here $\tau_0$ defines the reference complex structure on the curve
$$
T^2_{0}=\{0<x\leq 1,~0<y\leq 1, ~z=x+\tau_0y,~\bz=x+\bar{\tau}_0y\}.
$$

  For small $t_\tau$  from (\ref{2.6})
\beq{2b.6}
 \mu_\tau=
\ti{\mu}_\tau\bp(\bz-z)(1-\sum_{a=1}^n\chi_a'(z,\bz)),~~
(\ti{\mu}_\tau=\frac{t_\tau}{\tau-\tau_0}),
\eq
or
\beq{2a.6}
\mu_\tau=
\frac{t_\tau}{\rho}\bp(\bz-z)
(1-\sum_{a=1}^n\chi_a(z,\bz)).
\eq
As we assumed from the very
beginning, $\mu_\tau$ vanishes in the marked points.
It describes not only a small vicinity of $T^2_0$
in ${\cal M}_{1,1}$, but also the whole Teichm\"{u}ller space as well.
In terms of $\mu_\tau $ the Teichm\"{u}ller space  is the unity disk
$|\mu_\tau|<1$ (see (\ref{7a.2}) for the transformation law).
In terms of $\tau$ it  is the upper half plane Im$\tau >0$.
 On the other hand, the times
\mbox{${\bf t}=(t_\tau,t_1=x_1-x_1^0,\ldots,t_n=x_n-x_n^0)$}
define  deformations
$$T^2_{0}(x_1^0,\ldots,x_n^0)\stackrel{{\bf t}}
{\rar}T^2_\tau(x_1,\ldots,x_n)$$
in a vicinity of $T^2_{0}$ in ${\cal M}_{1,1}$.
Eventually, for small $t_\tau$ and $t_a$ we have from (\ref{2a.6})
\beq{2c.6}
\de\mu=\de \ti{\mu}_\tau\bp(\bz-z)(1-\sum_{a=1}^n\chi_a(z,\bz))+
 \sum_{a=1}^n\de t_a\bp(1+c_a(z-x_a))\chi_a(z,\bz))
\eq

\bigskip
{\bf 7.2. Flat bundles on a family of elliptic curves}

Note first, that $\bA$ as a $\p$-connection
determines a holomorphic $G$ bundle ${\cal E}$ over $T^2_{\tau}$.
For stable bundles $\bA$ can be gauge transformed by
(\ref{11.2}) to the Cartan $(z,\bz)$-independent form $\bar{L}$
$$\bA=f\bar{L}f^{-1}+f(\bp+\mu\p)f^{-1},$$
$$ \bar{L}\in{\cal H}-{\rm Cartan~subalgebra~of ~Lie}(G).$$
Therefore, a stable bundle ${\cal E}$ is decomposed into the direct
sum of line bundles
$${\cal E}=\oplus_{k=1}^r {\cal L}_k, ~~r=\mbox{rank}(G).$$
The set of gauge equivalent connections represented by $\{\bar{L}\}$
can be identified with the $r$ power of the Jacobian of $T^2_{\tau}$,
 factorized
by the action of the Weyl group $W$ of $G$.
Put
\beq{4.6}
\bar{L}=2\pi i \frac{1-\ti{\mu}_\tau}{\rho}{\bf u},
~~{\bf u}\in{\cal H},~~
(\frac{1-\ti{\mu}_\tau}{\rho}=\frac{1}{\tau-\bar{\tau}_0}).
\eq
It means that
\beq{4a.6}
\int_{T^2_\tau}\bar{L}dw={\bf u}.
\eq
Let $\bar{L}=\bp\log {\bf \phi}$. Then the integral
$$
\int_{T^2_\tau}\bar{L}dw=\int_{P_0}^P\log {\bf \phi}dw.
$$
defines the Abel map $P\in T^2_\tau\rar {\bf u}.$
We will come again to this point later.

The flatness condition (the moment constraints (\ref{13.2})) for
the gauge transformed connections $(L,\bar{L})$  takes the form
\beq{3.6}
\p_{\bar{w}}L+[\bar{L},L]=2\pi i\sum_{a=1}^n\de^2(x^0_a,)p_a.
\eq
Let $R=\{\al\}$ be the root system of of Lie$(G)={\cal G}$ and
$${\cal G}={\cal H}\oplus_{\al\in R}{\cal G}_\al$$
be the root decomposition.
Impose the vanishing of the residues in (\ref{3.6})
\beq{11.6}
\sum_{a=1}^np_a|_{\cal H}=0,
\eq
where $p_a|_{\cal H}$ is the Cartan component of $p_a$ and we have
identified ${\cal G}$ with its dual space ${\cal G}^*$.
This condition is similar to (\ref{VP}) and has the sense of the moment
constraints for the remnant gauge action (see below {\bf 7.3}).

We will parametrized the set of solutions of (\ref{3.6}) by two elements
${\bf v},{\bf u}\in {\cal H}$. Let
$E_1(w)$ be the Eisenstein function of module $\tau$ (\ref{A.1})
and $(p_a)_{\cal H}$ and $(p_a)_{\al}$
are the Cartan and the root component of $p_a\in{\cal O}_a$,
\\
\begin{lem}
Solutions of the moment constraint equation (\ref{3.6}) have the form
\beq{5.6}
L=P+X,~~P\in{\cal H},~~X=\sum_{\al\in R}X_{\al}.\eq
\beq{6.6}
P=2\pi i(\frac{{\bf v}}{1-\ti{\mu}_\tau}-\ka\frac{{\bf u}}{\rho}+
\sum_{a=1}^n(p_a)_{\cal H}E_1(w-x_a)),
\eq
\beq{7.6}
X_{\al}=\sum_{a=1}^nX_{\al}^a,
\eq
$$
X_{\al}^a=\frac{(p_a)_{\al}}{1-\ti{\mu}_\tau}\exp 2\pi i\{
\frac{(w-x_a)-(\bar{w}-\bar{x}_a)}{\tau-\bar{\tau}_0}
\al(\bfu)\}
\phi(\al(\bfu),w-x_a),
$$
where $\phi(u,w)$ is defined in (\ref{A.3}).
\end{lem}
{\sl Proof.} Consider first the Cartan  component. Since
$\bar{L}\in{\cal H}$
$$
\p_{\bar{w}}P=2\pi i\sum_{a=1}^n\de^2(x_a)p_a.
$$
From (\ref{G1}),(\ref{G2})  we obtain (\ref{6.6}).
The special choice of the constant part of $P$
will be explained later. 
Note, that $\bf{v}\in{\cal H}$ is a new parameter.

For the root components (\ref{3.6}) takes the form
$$
(\p_{\bar{w}}+2\pi i\frac{1-\ti{\mu}_\tau}{\rho}\al({\bf u}))X_{\al}=
2\pi i\sum_{a=1}^n\de^2(x_a,)(p_a)_\al.
$$
Comparing it with (\ref{G3}) and its solution
(\ref{G4}) we come to (\ref{7.6}).$\Box$

Therefore we have found the flat connections 
$L=L(\bfv,\bfu),\bar{L}=\bar{L}(\bfu).$

\bigskip
{\bf 7.3. Symmetries.}

The remnant gauge transforms do not change the gauge fixing and
thereby preserve the chosen Cartan subalgebra
${\cal H}\subset G$. These transformations are generated by the Weyl
 subgroup $W$ of $G$ and  elements
\mbox{$f(w,\bar{w})\in {\rm Map}(T^2_{\tau},{\rm Cartan}(G))$}.
Let $\Pi$ be the system of simple roots,
$R^{\vee}=\{\al^{\vee}=\frac{2\al}{(\al|\al)}\},$ is the dual
 root system,
and ${\bf m}=\sum_{\al\in \Pi} m_{\al}\al^{\vee}$
 be the element from the dual root
lattice ${\bf Z}R^{\vee}$. Then the Cartan valued harmonics
\beq{9.6}
f_{{\bf m},{\bf n}}=\exp 2\pi i
({\bf m}\frac{w-\bar{w}}{\tau-\bar{\tau}_0}+
{\bf n}\frac{\tau\bar{w}
-\bar{\tau}_0w}{\tau-\bar{\tau}_0}),~~
({\bf m} ,{\bf n}\in R^{\vee})
\eq
generate the basis in the space of gauge transforms.
They act as
$$
\bar{L}\rar \bar{L}+
2\pi i\frac{{\bf m}-{\bf n}\tau}{\tau-\bar{\tau}_0}.
$$
$$
P\rar P+2\pi i\ka\frac{-{\bf m}+{\bf n}\bar{\tau}_0}{\tau-\bar{\tau}_0},
$$
$$
X_{\al}^a\rar X_{\al}^a\vf(m_\al,n_\al),
$$
\beq{9a.6}
\vf(m_\al,n_\al)=
\exp\frac{4\pi i}{\tau-\bar{\tau}_0}[(m_\al+n_\al\bar{\tau}_0)(w-x_a)
-(m_\al+n_\al\tau)(\bar{w}-\bar{x}_a)].
\eq
In terms of the new variables ${\bf v}$ and ${\bf u}$
 they take especial simple form
\beq{10.6}
{\bf u}\rar {\bf u} +{\bf m}-{\bf n}\tau,~~ {\bf v}\rar  
{\bf v}-\ka{\bf n}.
\eq
The whole discrete gauge symmetry 
is the semidirect product $\hat{W}$ of
 the Weyl group
$W$ and the lattice ${\bf Z}R^{\vee}\oplus\tau{\bf Z}R^{\vee}$. 
It is the
Bernstein-Schvartsman complex crystallographic group \cite{BS}.
The factor space ${\cal H}/\hat{W}$ is  the genuine space
for the coordinates ${\bf u}$,
that we discussed above (see (\ref{4.6}) and (\ref{4a.6})).

The transformations (\ref{9.6}) according with (\ref{8.2}) act also on
$p_a\in{\cal O}_a$. This action leads to the symplectic quotient
${\cal O}_a//H$ and generates the moment equation (\ref{11.6}).

The modular group ${\rm PSL}_2({\bf Z})$ 
s a subgroup of mapping class group
for the Teichm\"{u}ller space ${\cal T}_{1,n}$. We don't consider here
the action of the permutation of the marked points on dynamical variables
 $({\bf v},{\bf u},{\bf p},\tau,x_a)$.
 Due to (\ref{7a.2}) and
(\ref{2b.6}) its action on $\tau$ takes the standard form
$$
\tau\rar\ga\tau=\frac{a\tau+b}{c\tau+d},~~\ga\in{\rm PSL}_2({\bf Z}).
$$
We summarize the action of the
Bernstein-Schvartsmann  group and the modular group on the dynamical
variables:
\bigskip
\begin{center}
\begin{tabular}{|c|c|c|c|}\hline
             &W=\{s\}&${\bf Z}R^{\vee}\oplus\tau{\bf Z}R^{\vee}$
&${\rm PSL}_2({\bf Z})$
\\ \hline \hline
${\bf v}$ &$s{\bf v}$& ${\bf v}+\ka{\bf n} $
& ${\bf v}(c\tau+d)-\ka c{\bf u} $
\\ \hline
${\bf u}$ &$s{\bf u}$  & ${\bf u}-{\bf m}+{\bf n}\tau$ 
 &${\bf u}(c\tau+d)^{-1}$
\\ \hline
$(p_a)_{\cal H}$ & $s(p_a)_{\cal H}$   &   $p_a$                        
 & $p_a$
\\   \hline
$(p_a)_{\al}$ & $(p_a)_{s\al}$   &   $\vf(m_\al,n_\al)(p_a)_{\al}$  
   & $(p_a)_\al$
\\   \hline
$\tau$   &$\tau$  &$\tau$                   &$\frac{a\tau+b}{c\tau+d}$
\\ \hline
$x_a$     &  $x_a$  &$x_a$        &    $\frac {x_a}{c\tau+d}$
\\ \hline
\end{tabular}
\end{center}
Here $\vf(m_\al,n_\al)$ is defined by (\ref{9a.6}).

\bigskip
{\bf 7.4 Symplectic form.}

The set $({\bf v},{\bf u})\in {\cal H},{\bf p}=(p_1,\ldots,p_n)
\in\oplus_{a=1}^n{\cal O}_a$
of dynamical variables along with the times 
${\bf t}=(t_\tau,t_1,\ldots,t_n)$
describe local coordinates in the total space of the bundle ${\cal P}$. 
According with the general
prescription, we can define the hamiltonian system on this set.
The main statement, formulated in terms of the theta-functions and the
Eisenstein functions (see Appendix), takes the form
\begin{predl}
The symplectic form $\om$ (\ref{19.2}) on ${\cal P}$ is
\beq{16.6}
\frac{1}{4\pi^2}\om=(\de{\bf v},\de{\bf u})+
\sum_{a=1}^n\de<p_a,g_a^{-1}\de g_a>
-\frac{1}{\ka}(\sum_{a=1}^n\de H_{2,a}+\de H_{1,a})\de t_a-
\frac{1}{\ka}\de H_{\tau}\de\tau,
\eq
with the Hamiltonians
\beq{14.6}
H_{2,a}=c_a<p_a,p_a>;
\eq
\beq{15.6}
H_{1,a}=
=2(\frac{{\bf v}}{1-\ti{\mu}_\tau}-
\ka\frac{\bfu}{\rho},p_a|_{\cal H})+
\sum_{b\neq a}(p_a|_{\cal H},p_b|_{\cal H})E_1(x_a-x_b)+
\eq
$$
\sum_{b\neq a}\sum_\al(p_a|_{\al},p_b|_{-\al})
\phi(\al({\bf u}),x_a-x_b);
$$
\beq{hamt}
H_{\tau}=\frac{({\bf v},{\bf v})}{2}-
\eq
$$\f1{4\pi^2}\sum_{a=1}^n
\sum_{\al}(p_a|_{\al},p_a|_{-\al})E_2(\al({\bf u}))+
\sum_{a\neq b}^n(p_a|_{\cal H},p_b|_{\cal H})(E_2(x_a-x_b)-
E_1^2(x_a-x_b))-
$$
$$
\f1{4\pi^2}\sum_{a\neq b}\sum_\al(p_a|_{\al},p_b|_{-\al})
\phi(-\al({\bf u}),x_a-x_b)
(E_1(\al({\bf u}))-E_1(x_b-x_a+\al({\bf u})));
$$
$$
\phi(\al({\bf u}),x_a-x_b)=
\frac
{\te(\al({\bf u})+x_a-x_b)\te'(0)}
{\te(\al({\bf u}))\te(x_a-x_b)}.
$$
\end{predl}
{\sl Proof.}
The form we have to calculate is (\ref{16.2})
$$
\om=\int_{\Si}<\de L,\de\bar{L}>+
2\pi i\sum_{a=1}^n\de<p_a,g_a^{-1}\de g_a>-
\f1{\ka}\int_{\Si}< L,\de L>\de\mu
$$
with $\de\mu$ (\ref{2c.6}).
First, we gauge transform $(L,\bar{L})$ by
$$
f(w,\bar{w})=\prod_{a=1}^n\exp(2\pi i
\frac{w-\bar{w}}{\tau-\tau_0}
\chi'_a(w,\bar{w}){\bf u}),
$$
where we choose $\chi'_a(w,\bar{w})$ in a such way that
\beq{sup}
{\rm supp}\chi'_a(w,\bar{w})\subset{\rm supp}\chi_a(w,\bar{w}),
\eq
$$
{\rm supp}\chi'_a(w,\bar{w})\cap{\rm supp}\bp\chi_a(w,\bar{w})
=\emptyset,
$$
and $\chi_a(w,\bar{w})$
 related to the moduli curves (\ref{2.6}). In fact, the first condition
follows from the second.

As we know the gauge transformations do not change $\om$, 
but do change relations between its summands.
Instead of (\ref{4.6}).(\ref{6.6}) and (\ref{7.6}) we obtain
\beq{bL}
\bar{L}=2\pi i \frac{1-\ti{\mu}_\tau}{\rho}{\bf u}\p_{\bar{w}}
(\bar{w}-w)(1-\sum_{a=1}^n\chi'_a(w,\bar{w})),
\eq
$$
P=2\pi i(\frac{{\bf v}}{1-\ti{\mu}_\tau}-\ka\frac{{\bf u}}{\rho}+
\sum_{a=1}^n(p_a)_{\cal H}E_1(w-x_a))-
$$
\beq{P}
-\frac{1-\ti{\mu}_\tau}{\rho}{\bf u}\p_{w}
(\bar{w}-w)\sum_{a=1}^n\chi'_a(w,\bar{w}),
\eq
\beq{7.6a}
X_{\al}=\f1{1-\ti{\mu}_\tau}
\exp\{\frac{w-\bar{w}}{\tau-\bar{\tau}_0}
\al(\bfu)\}
\sum_{a=1}^n(p_a)_{\al}\phi(\al(\bfu),w-x_a),
\eq

 Taking into account
the explicit form of $\bar {L}$ (\ref{4.6})  we obtain
\beq{12.6}
<\de L,\de{\bar L}>=
\frac{(\de{\bf v},\de{\bf u})}{\rho}+S(\de\tau,\de\bft),
\eq
where $S(\de\tau,\de\bft)$ is a sum of terms 
with a linear dependence on
the "time" differentials. It is compensated by terms coming from
$$
-\frac{1}{4\pi^2\ka}\int_{T^2_\tau}<L,\de L>\de\mu.
$$
Let us calculate the Hamiltonians
$$
-\frac{1}{4\pi ^2}<L, L>=
(\frac{{\bf v}}{1-\ti{\mu}_\tau}-
\ka\frac{{\bf u}}{\rho}+
(\sum_{a=1}^np_a|_{\cal H}E_1(w-x_a))^2-
$$
\beq{LL}
-\frac{4\pi^2}{1-\ti{\mu}_\tau}\sum_{a,b}\sum_{\al\in R}
(p_a|_{\al},p_b|_{-\al})
\phi(-\al({\bf u}),w-x_a)
\phi(\al({\bf u}),w-x_b)-
\eq
$$
\f1{2\pi i}(\frac{{\bf v}}{1-\ti{\mu}_\tau}-\ka\frac{{\bf u}}{\rho}+
\sum_{a=1}^n(p_a)_{\cal H}E_1(w-x_a)),
\frac{1-\ti{\mu}_\tau}{\rho}{\bf u})
\p_{w}
(\bar{w}-w)\sum_{a=1}^n\chi'_a(w,\bar{w}).
$$
To take the integral over $T^2_0$ we have to couple this two form with
$\de\mu$ (\ref{2c.6}) 
$$
-\frac{1}{4\pi^2}\int_{T^2_\tau}<L,L>\de\mu=
$$
$$
-\frac{1}{4\pi^2}\int_{T^2_\tau}<L,L>
\de \ti{\mu}_\tau\bp(\bar{w}-w)(1-\sum_{a=1}^n\chi_a(w,\bar{w}))+
 \sum_{a=1}^n\de t_a\p_{\bar w}(1+c_a(w-x_a))\chi_a(w,\bar w)).
$$
Due to our choice of $\chi'_a(w,\bar{w})$ (\ref{sup}),
the last line of (\ref{LL}) does not contribute in the integral. 
Therefore,
we leave with the holomorphic double periodic part of two-form $<L,L>$. 
Using  (\ref{A.7a})  we rewrite (\ref{LL}) as
$$
-\frac{1}{4\pi ^2}<L, L>=
(\frac{{\bf v}}{1-\ti{\mu}_\tau}-
\ka\frac{{\bf u}}{\rho},
\frac{{\bf v}}{1-\ti{\mu}_\tau}-
\ka\frac{{\bf u}}{\rho})+
$$
$$
2\sum_{a=1}^n(\frac{{\bf v}}{1-\ti{\mu}_\tau}-
\ka\frac{{\bf u}}{\rho},p_a|_{\cal H})E_1(w-x_a)+
$$
$$
\sum_{a\neq b}^n(p_a|_ {\cal H},p_b|_ {\cal H}) (E_1(w-x_a)-
E_1(w-x_b))-
$$
$$
-\frac{4\pi^2}{1-\ti{\mu}_\tau}\sum_{a\neq b}\sum_{\al\in R}
(p_a|_{\al},p_b|_{-\al})
\phi(-\al({\bf u}),w-x_a)
\phi(\al({\bf u}),w-x_b)-
$$
$$
-\frac{4\pi^2}{1-\ti{\mu}_\tau}\sum_{a}\sum_{\al\in R}
(E_2(w-x_a)-E_2(\al({\bf u})).
$$
Since $L$ has only first order poles (\ref{P}),(\ref{7.6a}), we expand
it on the deformed torus
 according with(\ref{A.17})
\beq{13.6}
-\frac{1}{4\pi^2}<L,L>=
(\sum_{a=1}^nH_{2,a}E_2(w-x_a)+H_{1,a}E_1(w-x_a))+h_0.
\eq
Due to (\ref{A.17}),(\ref{A.17b}) and (\ref{A.17a})
$$
-\frac{1}{4\pi^2}\int_{T^2_\tau}<L,L>\de\mu=
$$
$$
\sum_{a=1}^n(\int_{T^2_\tau}(H_{2,a}E_2(w-x_a)\p_{\bar w}(w-x_a)
\chi_a(w,\bar{w})+
H_{1,a}E_1(w-x_a))\p_{\bar w}\chi_a(w,\bar{w}))\de t_a
$$
$$
+h_0\de \ti{\mu}_\tau\bp(\bar{w}-w)(1-\sum_{a=1}^n\chi_a(w,\bar{w})).
$$
Taking into account and (\ref{A.18}),(\ref{A.19}) we find
$$
H_{2,a}=res|_{x_a}<L,L>(w-x_a)=
c_a<p_a,p_a>+const.,
$$
$$
H_{1,a}=res|_{x_a}<L,L>=
\mbox{(\ref{15.6})}.
$$
The constant term in (\ref{13.6}) $h_0$ is defined by (\ref{A.20}).
To find it we use (\ref{A.7a}) and (\ref{A.21}).
Then
$H_{\tau}=h_0\p_\tau\ti{\mu}_\tau$. After some algebra we obtain 
(\ref{hamt}).
$\Box$
\bigskip

{\bf 7.5 Example 1. $PVI_{\frac{\nu^2}{4},-\frac{\nu^2}{4},
\frac{\nu^2}{4},\f1{2}-\frac{\nu^2}{4}}$.}

Consider ${\rm SL}(2,{\bf C})$ bundles over the family of $\Si_{1,1}$.
Then (\ref{4.6}) takes the form
\beq{BL}
\bar{L}=2\pi i \frac{1-\ti{\mu}_\tau}{\rho}\di (u,-u).
\eq
In this case the position of the marked point is no long the module and we
put $x_1=0$. We have from (\ref{2b.6})
$$w=z-\frac{\tau-\tau_0}{\rho}(\bz-z),~\bar{w}=\bz,$$
$$\p_{\bar w}=\bp+\frac{\tau-\tau_0}{\tau-\bar{\tau}_0}\p.$$

Since $\dim{\cal O}=2$ the orbit degrees of freedom
can be gauged away by the hamiltonian action of the diagonal group.
We assume that
$$
p=\nu[(1,1)^T\otimes(1,1)-Id].
$$
Then we have from(\ref{5.6}),(\ref{6.6}),(\ref{7.6})
\beq{17.6}
L=\mat{2\pi i (\frac{v}{1-\ti{\mu}_\tau}-\ka\frac{u}{\rho})}
{x(2u,w,\bar{w})}{x(-2u,w,\bar{w})}
{2\pi i (-\frac{v}{1-\ti{\mu}_\tau}+\ka\frac{u}{\rho})}.
\eq
$$
x(u,w,\bar{w})=\frac{\nu}{2\pi i(1-\ti{\mu)}_\tau}
\exp 2\pi i\{(w-\bar{w})u
\frac{1-\ti{\mu}_{\tau}}{\rho}\}\phi(u,w).
$$

The symplectic form (\ref{16.6})
$$
-\frac{1}{8\pi^2}\om=(\de v,\de u)-\frac{1}{\ka}\de H_{\tau}\de\tau,
$$
and
$$
H_{\tau}=\frac{v^2}{2}+U(u|\tau),~~U(u|\tau)=
-(\frac{\nu}{2\pi i})^2E_2(2u|\tau)-\frac{\nu^2}{2\pi i}.
$$
Then the equations of motion are
\beq{20.6}
\ka\frac{\p u}{\p\tau}=v,
\eq
\beq{21.6}
\ka\frac{\p v}{\p\tau}=\frac{\nu^2}{(2\pi i)^2}\frac{\p}{\p u}E_2(2u|\tau).
\eq
Due to (\ref{A.7a}) we obtain
$$
\ka^2\frac{\p^2 u}{\p\tau^2}=
\frac{\nu^2}{(2\pi i)^2}\frac{\p}{\p u}\wp(2u|\tau).
$$
which coincides with (\ref{I.8}).
This equation provides the isomonodromic deformation
for the linear system
(\ref{23.2}),(\ref{25a.2}) with $L$ (\ref{17.6}) and $\bar{L}$ (\ref{BL})
with respect to change the module $\tau$.
The Lax pair is given by $L$ (\ref{17.6}) and $M_\tau$
$$
M_\tau=\mat{0}{y(2u,w,\bar{w})}{y(-2u,w,\bar{w})}{0},
$$
where $y(u,w,\bar{w})$ is defined  by the equation (see (\ref{25.2})
\beq{y}
(\p_{\bar{w}}+\frac{2\pi i}{\tau-\bar{\tau}_0}u)y(u,w,\bar{w})=
-\frac{\rho}{\ka(\tau-\bar{\tau}_0)^2}x(u,w,\bar{w}).
\eq
Using  the representation (\ref{G6}) for $x(u,w,\bar{w})=g_2(u,w,\bar{w})$
we find
\beq{22.6}
y(u,w,\bar{w})=
\frac{\rho}{2\pi i\ka(\tau-\bar{\tau}_0)}\p_u x(u,w,\bar{w}).
\eq
The equivalence of the Lax equation
$$\p_\tau L-\ka\p_w M+[M,L]=0$$
to the equations of motion (\ref{20.6}),(\ref{21.6}) can be checked
by substituting $L$ (\ref{17.6}) and $M$ with $y(u,w,\bar{w})$`
(\ref{22.6}).
Details of this procedure will be replenished  for the $\SL$ bundles
in the next example.

The projection method determines  solutions of (\ref{20.6}),(\ref{21.6})
 as a result of
diagonalization of $L$ (\ref{17.6}) by the gauge transform on the deformed
curve $T^2_\tau$
$$
\f1{\tau-\bar{\tau}_0}\di (u(\tau),-u(\tau))=
$$
$$
f(z,\bz,\tau)(
\frac{2\pi i (\tau-\tau_0)}{\rho}
\mat{v^0-\ka\frac{u^0}{\rho}}
{x(2u^0,z,\bar{z})}{x(-2u^0,z,\bar{z})}
{-v^0+\ka\frac{u^0}{\rho}}+
\f1{\rho}\di (u^0,-u^0))
f^{-1}(z,\bz,\tau)+
$$
$$
+f^{-1}(z,\bar{z},\tau)(\bp +\frac{\tau-\tau_0}{\tau-\bar{\tau}_0}\p)
f(z,\bz,\tau).
$$
On the critical level $(\ka=0)$ we come to the two-body elliptic Calogero
 system.

{\bf 7.6. Example 2.}

For  flat $G$ bundles over $\Si_{1,1}$ we obtain PVI-type equations,
related to arbitrary root systems. They are described by the system
of second order  differential equations for the
${\bf u}=(u_1,\ldots,u_r),~(r=$rank$G)$ variables.
In addition, there are the orbit variables $\bfp\in{\cal O}(G)$
satisfying the Euler top equations.
Consider in detail the $\SL$ case with  the most degenerate orbits
${\cal O}=T^*{\bf C}P^{N-1}$. They have dimension
$2N-2$. The orbit variables can be gauged away by the diagonal 
gauge transforms and we are left with the coupling constant $\nu$.
We already have the $\bar {L}$ and $L$ matrices
(see (\ref{4.6}) and (\ref{5.6}))
$$
L=P+X,
$$
$$
P=2\pi i(\frac{{\bf v}}{1-\ti{\mu}_\tau}-\ka\frac{{\bf u}}{\rho}),
$$
$$
X=\{x_{\al}\}=(\tau-\bar{\tau}_0)\nu\exp 2\pi i\{
\frac{w-\bar{w}}{\tau-\bar{\tau}_0}
\al(\bfu)\}
\phi(\al(\bfu),w).
$$
Here $\al=e_j-e_k, ~\al(\bfu)=u_j-u_k,~(j\neq k),$ and
$${\bf u}=\di(u_1,\ldots,u_N),~~{\bf v}=\di(v_1,\ldots,v_N).$$
The equations of motion take the form
\beq{eq1}
\ka \frac{du_j}{d\tau}=v_j,
\eq
\beq{eq2}
\ka \frac{dv_j}{d\tau}=-\p_{u_j} U(\bfu|\tau),
\eq
$$
U(\bfu|\tau)=\f1{(2\pi i)^2}\sum_{j\neq k}p_{j,k}p_{k,j}
E_2(u_j-u_k|\tau).
$$
On the critical level this Painlev\'{e} type
system degenerates into $N$-body elliptic Calogero
system.

Let us check the consistency of the equations of motion
with the Lax representation
\beq{La}
\p_\tau L-\ka\p_wM+[M,L]=0.
\eq
 Put the $M$ operator  in the form
$$M=-D+Y,$$
where
$$Y= \{y(u_j-u_k)\}=\{y_\al\},$$
and $D$ is a diagonal matrix. We have already found $y_\al$ (\ref{22.6})
using (\ref{25.2}).
But the diagonal part $D$ is not fixed by (\ref{25.2}) 
and should be found from
the consistency of the Lax equation with the equations of motion.
 We take it in the same form as in the Calogero system \cite{Ca1}
$$D=\di(d_1,\ldots,d_N),~~d_j=\sum_{i\neq j}^Ns(u_j-u_i).$$

First, we will prove few facts concerning the matrix elements 
of $L$ and $M$.
We will prove that they satisfy the following functional equation
\beq{sf}
x(u,z,\bz)y(v,z,\bz)-x(v,z,\bz)y(u,z,\bz)=(s(v)-s(u))x(u+v,z,\bz).
\eq
In particular, its solution $s(u,w)$ is $w$ independent
\beq{ff}
s(u)=\f1{\ka}\wp(u)+const.
\eq
The  relation (\ref{sf}) is the so called Calogero functional equation.
Due to (\ref{22.6}) we can put in it the derivatives of $x$ instead
of $y$.
The exponential factor in the expression of $x$  cancels and
(\ref{sf}) take the form of the addition formula (\ref{ad2}). 
Simultaneously, we obtain (\ref{ff}).
We also need the following identity
\beq{23.6}
(\p_\tau-\frac{\rho}{2\pi i}\p_z\p_u+\frac{u}{\tau-\bar{\tau}_0}\p_u)
x(u,z,\bar{z})=0.
\eq
It can be derived from the representation  (\ref{G6})
for $x(u,z,\bz)=g_2(u,z,\bz)$.

Consider now the Lax equation (\ref{La}).
Separation the diagonal and nondiagonal terms  leads to the system
\beq{first}
 \frac{d}{d\tau}P+\ka\p_wD+\sum_\al(y_\al x_{-\al}-y_{-\al}x_{\al})=0,
\eq
\beq{sec}
\frac{d}{d\tau}x_{\al}-\ka\p_wy_\al-\al(D)x_{\al}+[Y,X]_\al=0.
\eq
 Due to (\ref{ff}) and (\ref{eq1}) the first equation  (\ref{first})
can be rewritten as
$$
2\pi i \frac{d}{d\tau}v_k=\frac{(1-\ti{\mu}_\tau)^2}{2\pi i\ka}
\sum_{j\neq k}(x(u_j-u_k)x'(u_k-u_j)-x(u_k-u_j)x'(u_j-u_k))=
$$
$$
\sum_{j\neq k}[x(u_j-u_k)x(u_k-u_j)]'.
$$
But
$$
x(u)x(-u)=\frac{\nu^2}{(1-\mu)^2}\phi(u,w)\phi(-u,w)=
\frac{\nu^2}{(1-\mu)^2}(E_2(w)-E_2(u)),
$$
(see (\ref{A.7a})). Thereby we come to the (\ref{eq2}).

Now check the off-diagonal part (\ref{sec}). It is convenient 
to go back from
$$
w=(1+\frac{\tau-\tau_0}{\rho})z-\frac{\tau-\tau_0}{\rho}\bz,~~
\bar{w}=\bz
$$
to $(z,\bz)$ variables. Then (\ref{sec}) takes the form
$$
\frac{d}{d\tau}x(u_j-u_k)-\ka\p_zy(u_j-u_k)-
2\pi i(\frac{v_j-v_k}{1-\ti{\mu}_\tau}-
\ka\frac{u_j-u_k}{\rho})y(u_j-u_k)-
$$
$$
-x(u_j-u_k)[\sum_{i\neq j}^Ns(u_j-u_i)-\sum_{i\neq k}^Ns(u_k-u_i)]+
$$
$$
+\sum_{i=1}^Ny(u_j-u_i)x(u_i-u_k)-y(u_i-u_k)x(u_j-u_i)=0.
$$
It follows from (\ref{22.6}),(\ref{eq1}), and (\ref{23.6}) that
$$
\frac{d}{d\tau}x(u_j-u_k)-\ka\p_zy(u_j-u_k)-2\pi i
(\frac{v_j-v_k}{1-\ti{\mu}_\tau}-\ka\frac{u_j-u_k}{\rho})y(u_j-u_k)=0.
$$
On the other hand the equality
$$
x(u_j-u_k)(s(u_j-u_i)-s(u_k-u_i))+y(u_j-u_i)x(u_i-u_k)-
y(u_i-u_k)x(u_j-u_i)=0.
$$
is the addition formula (\ref{sf}). It concludes the proof
of the equivalence of the Lax equation and the equations of motion.

The new ingredients of this construction in compare with the original
form \cite{Ca1} are the dependence the $L$-matrix on the spectral
parameter\footnote
{The dependence on the spectral parameter first introduced 
in \cite{Kr2} for the Calogero system in a slightly different form.}
, the presence of derivative $\p_zM$ in the Lax equation,
and the replacement of
the external time $t$ on the modular parameter $\tau$. Nevertheless,
the form of the Lax matrices is
defined as for the Calogero system by the same functional equation 
(\ref{sf}).
It turns out
that its solutions satisfy additional differential equations (\ref{ff}),
(\ref{23.6}), which allows to apply them in the isomonodromic
situation as well. Again, the solutions of the equations 
of motion (\ref{eq1}), (\ref{eq2}) are obtained by the diagonalization
$$
 2\pi i \frac{1-\ti{\mu}_\tau}{\rho}\di (u_1,\ldots,u_N)=\ti{\mu}_\tau
f^{-1}(L(\bfv^0,\bfu^0,\tau^0) f+
$$
$$
f^{-1}(\bp+\ti{\mu}_\tau\p)f.
$$

\section*{Appendix A}
\setcounter{equation}{0}
\def\theequation{A.\arabic{equation}}

We summarize the main formulae for elliptic functions, borrowed
mainly from \cite{We} .
We assume that $q=\exp 2\pi i\tau$, and the curve 
$T^2_\tau$ is $\bf{C}/\bf{Z}+\tau\bf{Z}$
factor of ${\bf C}$ under the shifts generated by $(1,\tau)$.

  The basic element is the theta  function:
\beq{A.1a}
\te(z|\tau)=q^{\frac
{1}{8}}\sum_{n\in {\bf Z}}(-1)^ne^{\pi i(n(n+1)\tau+2nz)}=
\eq
$$
q^{\frac{1}{8}}e^{-\frac{i\pi}{4}} (e^{i\pi z}-e^{-i\pi z})
\prod_{n=1}^\infty(1-q^n)(1-q^ne^{2i\pi z})(1-q^ne^{-2i\pi z})
 $$
{\sl The  Eisenstein functions.}
\beq{A.1}
E_1(z|\tau)=\p_z\log\te(z|\tau), ~~E_1(z|\tau)\sim\f1{z}+\ldots,
\eq
\beq{A.2}
E_2(z|\tau)=-\p_zE_1(z|\tau)=
\p_z^2\log\te(z|\tau),
~~E_2(z|\tau)\sim\f1{z^2}+\ldots.
\eq

The next important function is
\beq{A.3}
\phi(u,z)=
\frac
{\te(u+z)\te'(0)}
{\te(u)\te(z)}.
\eq
It has a pole at $z=0$ and
\beq{A.3a}
res|_{z=0}\phi(u,z)=1.
\eq

{\sl Relations to the Weierstrass functions. }
\beq{A.4}
\ze(z|\tau)=E_1(z|\tau)+2\eta_1(\tau)z,
\eq
\beq{A.5}
\wp(z|\tau)=E_2(z|\tau)-2\eta_1(\tau),
\eq
where
\beq{A.6}
\eta_1(\tau)=\ze(\frac{1}{2})=
\eq
$$
\frac{3}{\pi^2}\sum_{m=-\infty}^{\infty}\sum_{n=-\infty}^{\infty '}
\frac{1}{(m\tau+n)^2}=\frac{24}{2\pi i}\frac{\eta'(\tau)}{\eta(\tau)},
$$
where
$$
\eta(\tau)=q^{\frac{1}{24}}\prod_{n>0}(1-q^n).
$$
is the Dedekind function.

\beq{A.7}
\phi(u,z)=\exp(-2\eta_1uz)
\frac
{\si(u+z)}{\si(u)\si(z)}.
\eq
\beq{A.7a}
\phi(u,z)\phi(-u,z)=\wp(z)-\wp(u)=E_2(z)-E_2(u).
\eq

{\sl Series representations}
\beq{A.8}
E_1(z|\tau)=-2\pi i(\frac{1}{2}+\sum_{n\neq 0}
\frac{e^{2\pi iz}}{1-q^n})=
\eq
$$
-2\pi i(\sum_{n<0}\frac{1}{1-q^ne^{2\pi iz}}+
\sum_{n\geq 0}\frac{q^ne^{2\pi iz}}{1-q^ne^{2\pi iz}}+\frac{1}{2}).
$$
\beq{A.9}
E_2(z|\tau)
=-4\pi^2\sum_{n\in{\bf Z}}\frac{q^ne^{2\pi iz}}{(1-q^ne^{2\pi iz})^2}.
\eq
\beq{A.10}
\phi(u,z)=2\pi i\sum_{n\in{\bf Z}}\frac
{e^{-2\pi inz}}{1-q^ne^{-2\pi iu}}.
\eq

{\sl Parity.}\\
\beq{P.1}
\te(-z)=-\te(z)
\eq
\beq{P.2}
E_1(-z)=-E_1(z)
\eq
\beq{P.3}
E_2(-z)=E_2(z)
\eq
\beq{P.4}
\phi(u,z)=\phi(z,u)=-\phi(-u,-z)
\eq

{\sl Behaviour on the lattice}
\beq{A.11}
\te(z+1)=-\te(z),~~~\te(z+\tau)=-q^{-\oh}e^{-2\pi iz}\te(z),
\eq
\beq{A.12}
E_1(z+1)=E_1(z),~~~E_1(z+\tau)=E_1(z)-2\pi i,
\eq
\beq{A.13}
E_2(z+1)=E_2(z),~~~E_2(z+\tau)=E_2(z),
\eq
\beq{A.14}
\phi(u+1,z)=\phi(u,z),~~~\phi(u+\tau,z)=e^{-2\pi iz}\phi(u,z).
\eq

{\sl Modular properties}
\beq{A.15}
\te(\frac{z}{c\tau+d}|\frac{a\tau+b}{c\tau+d})=
\ep e^{\frac{\pi i}{4}}(c\tau+d)^\oh
\exp(\frac{i\pi cz^2}{c\tau+d})\te(z|\tau),~~(\ep^8=1).
\eq
\beq{A.15a}
E_1(\frac{z}{c\tau+d}|\frac{a\tau+b}{c\tau+d})=(c\tau+d)E_1(z|\tau)
+2\pi i z.
\eq
\beq{A.16}
E_2(\frac{z}{c\tau+d}|\frac{a\tau+b}{c\tau+d})=(c\tau+d)^2E_2(z|\tau)+
2\pi i(c\tau+d) .
\eq

{\sl Addition formula}\\
\beq{ad1}
\phi(u,z)\p_v\phi(v,z)-\phi(v,z)\p_u\phi(u,z)=(E_2(v)-E_2(u))\phi(u+v,z),
\eq
or
\beq{ad2}
\phi(u,z)\p_v\phi(v,z)-\phi(v,z)\p_u\phi(u,z)=(\wp(v)-\wp(u))\phi(u+v,z).
\eq
The proof of (\ref{ad1}) is based on (\ref{A.3a}),(\ref{P.4}), 
and (\ref{A.14}).

In fact, $\phi(u,z)$ satisfies more general relation which follows from the
Fay three-section formula
\beq{ad3}
\phi(u_1,z_1)\phi(u_2,z_2)-\phi(u_1+u_2,z_1)\phi(u_2,z_2-z_1)-
\phi(u_1+u_2,z_2)\phi(u_1,z_1-z_2)=0
\eq

{\sl Green functions}\\
The Green functions are $(1,0)$ forms on $T_\tau^2$.  The first
$g_1(z)$ is defined by the equation
 \beq{G1}
\bp g_1(z)=2\pi i\sum_ap_a\de^2(x_a), ~~(\sum_ap_a=0),
\eq
where
\beq{G5}
\de^2(x_a)=\sum_{m,n\in {\bf Z}}f_{m,n}(z-x_a,\bz-\bar{x}_a),
\eq
$$
f_{m,n}(z,\bz)=\exp\frac{2\pi i}{\rho}\{m(z-\bz)+
n(\tau\bz-\bar{\tau}z)\},
~~(\rho=\tau-\bar{\tau}),
$$
\beq{G2}
g_1(z)=\sum_ap_aE_1(z-x_a)+const.
\eq
For the equation
\beq{G3}
(\bp +\frac{2\pi i}{\rho}u)g_2(u,z)=2\pi i\de^2(0),
\eq
the Green function is
\beq{G4}
 g_2(u,z)=\f1{\rho}e^{2\pi i\frac{z-\bz}{\rho}u}\phi(u,z).
\eq
\beq{G6}
 g_2(u,z)=\rho\sum_{m,n\in {\bf Z}}\frac{f_{m,n}(z,\bz)}
{u-m+n\tau}
\eq

{\sl Expansion of elliptic functions.}\\
Let $M_2$ be the space of meromorphic elliptic function
on $T^2_0$ with poles of order two or less. Then any $f(z)\in M_2$
 can be decomposed in the sum of the Eisenstein functions
\beq{A.17}
f(z)=\sum_a (c_{2,a}E_2(z-x_a|\tau)+c_{1,a}E_1(z-x_a|\tau))+c_0, ~~
\eq
where
\beq{A.18}
\sum_a c_{1,a}=0,
\eq
\beq{A.19}
c_{1,a}=res|_{x_a}f(z),~~~c_{2,a}=res|_{x_a}(z-x_a)f(z),
\eq
and
\beq{A.20}
c_0=\mbox{const. part}
[f(z)-\sum_a (c_{2,a}E_2(z-x_a|\tau)+c_{1,a}E_1(z-x_a|\tau)].
\eq
In particular, for $a\neq b$
\beq{A.21}
\mbox{const. part}[\phi(u,z-x_a)\phi(-u,z-x_b)]=
\phi(-u,x_a-x_b)[E_1(u)-E_1(u+x_b-x_a)].
\eq
According with (\ref{A.17}) denote
\beq{A.17b}
e_{2,a}=E_2(z-x_a),~~e_{1,a}=E_1(z-x_a),~~e_0=1
\eq
the basis in $M_2$. The dual basis with respect to the integration on 
$T^2_0$ is
\beq{A.17a}
f_{2,a}=\bp(z-x_a)\chi_a(z,\bz),~~f_{1,a}=\bp\chi_a(z,\bz),~~
f_0=\bp(\bz-z)(1-\sum_{a=1}^n\chi_a(z,\bz)),
\eq
where $\chi_a$ is the characteristic function of vicinity ${\cal U}_a$ of
$x_a$ (see (\ref{cf})).

{\sl Integrals.}\\
\beq{A.28}
\sum_a c_{1,a}\int_{T^2_\tau}E_1(z-x_a|\tau)=\sum_a c_{1,a}(x_a-\bar{x}_a).
\eq
\beq{A.29}
\int_{T^2_\tau}E_2(z-x_a|\tau)=-2\pi i
\eq

\small{

}

\begin{thebibliography}{60}
\bibitem{FN}\ H.Flashka, A.Newell, Comm. Math. Phys. {\bf 76} (1980), 67
\bibitem{JMU}\ M. Jimbo, T. Miva and K. Ueno, {\em Monodromy preserving
deformations of linear ordinary differential equations}, I,II, Physica
 {\bf 2D} (1981), 306-352, 407-448
\bibitem{H1}\ N. Hitchin, {\em Stable  bundles and Integrable Systems},
Duke Math. Journ., {\bf 54} (1987) 91-114
\bibitem{KZ}\ V. Knizhnik and A. Zamolodchikov,
{\em Current algebra and Wess-Zumino model in two dimensions},
 Nucl.Phys. {\bf B247} (1984) 83
\bibitem{B}\ D. Bernard, {\em On the Wess-Zumino-Witten models on the torus},
Nucl.Phys. {\bf B303} (1988) 77; {\em On the Wess-Zumino-Witten models on the
Riemann surfaces},  Nucl.Phys. {\bf 309} (1988) 145
\bibitem{I1}\ D.Ivanov, {\em Knizhnik-Zamolodchikov-Bernard equations
on Riemann surfaces}, Int. J. Mod. Phys. {\bf A10} (1995) 2507-2536
\bibitem{Ne}\ N. Nekrasov, {\em Holomorphic bundles and many-body 
systems}, PUPT-1534, Comm. Math. Phys., {\bf 180} (1996) 587-604; 
 hep-th/9503157
\bibitem{ER}\  B. Enriques and V. Rubtsov,
{\em Hitchin systems, higher Gaudin operators
and r-matrices}, Math. Res. Lett. {\bf 3} (1996) 343-357
\bibitem{Fe}\ G.Felder, {\em \em The KZB equations
on Riemann surfaces}, hep-th/9609153
\bibitem{I2}\ D.Ivanov, {\em KZB eqs. as a quantization of nonstationary
Hitchin systems},  hep-th/9610207
\bibitem{BD}\ A.Beilinson, V.Drinfeld, {\em Quantization of Hitchin's
fibration and Langlands program}, preprint (1994)
\bibitem{FG}\ F.Falceto, K.Gawedzky, {\em Elliptic Wess-Zumino-Witten
model from elliptic Chern-simons theory,} hep-th/9502161,
Lett.Math.Phys. {\bf 38} (1996) 155
\bibitem{EK}\ P.Etingof, A.Kirillov, {\em Representations of affine
Lie algebras, parabolic differential equations and L\'{a}me functions},
Duke Math. J. {\bf 74} (1994), 585-614
\bibitem{FW}\ G.Felder,  C.Weiczerkovski, {\em Conformal blocks on 
elliptic curves and Knizhnik-Zamolodchikov-Bernard equations},
hep-th/9411004, Comm.Math.Phys. {\bf 176} (1996) 113
\bibitem{H2}\ N. Hitchin, {\em Flat connections and geometric quantization},
 Comm.Math.Phys., {\bf 131} (1990) 347-380
\bibitem{ADPW}\ S. Axelrod, S. Della Pietra  and E. Witten,
{\em Geometric quantization of the Chern-Simons gauge theory},
Journ. Diff. Geom., {\bf 33} (1991) 787-902
\bibitem{R}\ N. Reshetikhin, {\em The  Knizhnik-Zamolodchikov system as a
deformation of the isomonodromic problem},
 Lett. Math. Phys. {\bf 26} (1992) 167
\bibitem{Ha}\ J. Harnard, {\em Quantum isomonodromic deformations and the
Knizhnik-Zamolodchikov equations},  hep-th/9406078
\bibitem{Ko}\ D. Korotkin and J. Samtleben, {\em On the quantization
of isomonodromic deformations on the torus}, hep-th/9511087,
Int.J.Mod.Phys {\bf A12} (1997) 2013-2030
\bibitem{O4}\ K.Okamoto {\em D\'eforamation d'une \'equation 
differ\'entielle lin\'eare avec une singularit\'e
irreguli\'ere sur un tore}, J. Fac. Sci. Univ. Tokyo, Sect IA,
 {\bf 26} (1979) 501-518
\bibitem{Iw}\ K.Iwasaki {\em Fuchsian moduli on a Riemann surface - its 
Poisson structure and Poincar\'e-Lefschetz duality},
Pacific J. Math. {\bf 155} (1992) 319-340
\bibitem{PT}\ {\em Workshop on Painlev\'{e} Transcedents, 
Their asymptotics and Physical Applications}, NATO ASI Ser. B:
Physics Vol. 278 (1990) Sainte Adele,  
Quebec, ed. by D. Levi and P. Winternitz
\bibitem{O1}\ K.Okamoto, {\em Isomonodromic deformation 
and the  Painlev\'{e}
equations and the Garnier system}, J. Fac. Sci. Univ. Tokyo, Sect IA,
 Math. {\bf 33} (1986), 575-618
\bibitem{G}\ B. Gambier, {\em Sur les \'{e}quations differentielles du
 second ordre
et du premier degr\'{e} dont l'integral g\'{e}n\'{e}rale a ses points 
critiques
fixes}, Acta Math. Ann {\bf 33} (1910) 1-55
\bibitem{F}\ R. Fuchs, {\em \"{U}ber lineare homogene 
Differentialgleichungen zweiter
ordnung mit im endlich gelegne wesentlich singul\"{a}ren Stellen}, 
Math. Annalen,
{\bf 63} (1907) 301-323
\bibitem{Pa}\ Painlev\'{e}, CRAS, {\bf 143} (1906) 1111-1117
\bibitem{Ma}\ Yu.I.Manin, {\em Sixth Painlev\'{e} equation, universal
elliptic curve, and mirror of $P^2$}, Preprint Max-Planck-Institut (1996),
alg-geom/9605010
\bibitem{Ca}\ F.Calogero, {\em Exactly solvable one-dimensional
many-body problems}, Lett. Nuovo Cim. {\bf 13} (1976) 411-417
\bibitem{In}\ V.Inozemtsev, {\em Lax Representation with spectral
parameter on a torus for integrable particle systems}, Lett. Math. Phys.
{\bf 17} (1989) 11-17
\bibitem{TV}\ A.Treibich, J.-L.Verdier, {\em Rev\^{e}tements tangentiels 
et sommes de 4 nombres triangulaires}, C.R. Ac. Sci.
Paris, s\'{e}r. Math., {\bf 311} (1990), 51-54
\bibitem{H3}\ N.Hitchin, {\em Twistors spaces, Einstein metrics and
isomonodromic deformations}, Journ. Diff. Geom. {\bf 3} (1995) 52-134
\bibitem{O3}\  K.Okamoto, {\em Studies on the Painlev\'{e}
equations I, Sixth Painlev\'{e}}, Annali di Matematica Pura ed Applicata
{\bf 146} (1987) 337-381
\bibitem{Kr1}\ I. Krichever, {\em The tau-function of the universal Whitham
hierarchy, matrix models and topological field theories}, Comm. on Pure and
Appl. Math. {\bf XLVII} (1994) 437-475
\bibitem{O2}\  K.Okamoto, {\em On the tau-function of the Painlev\'{e}
equations}, Physica D {\bf 2} (1981) 525-535
\bibitem{IN}\ A.Its, V.Novokshenov, {\em The isomonodromic deformations 
method in the theory of Painlev\'e equations}, Lect. Notes in Math. 1191,
Springer-Verlag
\bibitem{Ta}\ K.Takasaki, {\em Spectral Curves and Whitham Equations in the
Isomonodromic Problems of Schlesinger Type} solv-int/9704004
\bibitem{VN}\ S.Novikov, A.Veselov, {\em On Poissson brackets compatible
with algebraic geometry and Korteweg-de Vries dynamics on the space of
finite-zone potentials}, Sov. Math. Doklady {\bf 26} (1982) 357-362
\bibitem{V}\ V.Vereshchagin, {\em Nonlinear quasiclassics and Painlev\'e
equations}, hep-th/9605092
\bibitem{Kr2}\ I.Krichever, {\em Elliptic solutions of 
the Kadomtsev-Petviashily
equation and many-body problems}, Funct. Anal. Aplic {\bf 14} (1980) 45
\bibitem{OP}\ M.Olshanetsky,  A.Perelomov , {\em Explicit solutions of the
Calogero models in the classical case and geodesic flows on symmetric spaces
of zero curvature},  Lett. Nuovo Cim. {\bf 16} (1976) 333-339;
{\em Explicit solutions of some completely integrable systems},
 Lett. Nuovo Cim. {\bf 17} (1976) 97-101
 \bibitem{Ca1}\ F.Calogero, {\em On a functional equation connected with
integrable many-body problems}, Lett. Nuovo Cim. {\bf 16} (1976) 77-80
Lectures, given at Spring school, Trieste (1990).
\bibitem{BS}\ J.Bernstein, O.Schvartsman, {\em Chevaley theorem
for complex crystallographic groups,} Funct. Anal. Appl. {\bf 12}
(1978), 308-310
\bibitem{We}\ Weyl A., {\em Elliptic functions according to Eisenstein and
Kronecker}, Springer, 1976
\end{thebibliography}
\end{document}